\documentclass[a4paper,11pt]{article}
\pdfoutput=1 
\usepackage{jcappub}
\usepackage{graphicx,amsmath,amsfonts,amssymb,slashed,float}
\usepackage{array,multirow}
\usepackage[utf8]{inputenc}
\usepackage[dvipsnames]{xcolor} 
\usepackage{dcolumn}
\usepackage{bbold}
\usepackage{bm}
\usepackage[normalem]{ulem}
\usepackage{mathrsfs} 
\usepackage{comment} 
\usepackage[T1]{fontenc} 
\usepackage{dirtytalk}
\usepackage{lmodern}
\usepackage{anyfontsize}
\usepackage{booktabs}
\usepackage{geometry}
\usepackage{mathtools}
\usepackage{relsize}
\newcommand{\pazocal}[1]{\mathcal{#1}}
\newcommand{\dr}{{\rm{dr}}} 
\newcommand{\nr}{{\rm{nr}}} 
 
\newcommand{\mH}{m_{\nu H}} 
\newcommand{\ml}{m_{\nu l}}

\newcommand{\Planck}{\textsl{Planck}}
\newcommand{\lcdm}{$\Lambda$CDM}

\newcommand{\cnb}{C$\nu$B}
\newcommand{\tU}{t_{\mathrm{U}}}

\renewcommand{\arraystretch}{1.2}

\title{\boldmath Impact of neutrino decays on the Cosmic Neutrino Background anisotropies}

\author[a]{Nicola Terzaghi,}
\author[a]{Guillermo Franco Abellán,}
\author[a]{Fabian Zimmer,}
\author[a,b]{Shin'ichiro Ando}

\affiliation[a]{GRAPPA Institute, University of Amsterdam, Science Park 904, 1098 XH Amsterdam, The
Netherlands}
\affiliation[b]{Kavli Institute for the Physics and Mathematics of the Universe, University of Tokyo, Chiba
277-8583, Japan}

\emailAdd{nicola.terzaghi@gmail.com}
\emailAdd{g.francoabellan@gmail.com}
\emailAdd{f.zimmer@uva.nl}
\emailAdd{s.ando@uva.nl}

\abstract{
The anisotropies of the Cosmic Neutrino Background (\cnb) offer an ideal tool to test non-standard neutrino interactions, since they directly trace the perturbations in the neutrino distribution function. Here, we study how invisible neutrino decays impact the  C$\nu$B anisotropies, in a framework where neutrinos decay non-relativistically to dark radiation and lighter neutrinos in a manner consistent with the measured mass splittings. For this purpose, we perform the first implementation of such a late-time neutrino decay scenario within a linear Einstein-Boltzmann solver, and compute the \cnb~angular power spectra from the Boltzmann hierarchy solutions for a range of lifetimes and decay channels. We find that neutrino decays leave very strong signatures on the \cnb~angular spectra, about two orders of magnitude larger than on the CMB angular spectra, particularly for lifetimes comparable to the age of the Universe. We show that a future polarized tritium target run of the PTOLEMY experiment, with sufficient counting statistics to measure just the first $\sim 15$ multipoles of the neutrino sky map, could test neutrino decay models that remain undetectable with CMB data.\\ 

\noindent\texttt{GitHub:} Our modified version of the Boltzmann code \texttt{CLASS} is publicly available \href{https://github.com/GuillermoFrancoAbellan/CLASSpp_nuDecay}{here}.

}

\begin{document}
\maketitle
\flushbottom

\section{Introduction}
\label{sec:intro}

More than six decades after the first detection of neutrinos, many of their fundamental properties remain unknown. Importantly, although oscillation experiments have firmly established the massive nature of neutrinos, these measurements only probe the mass-squared splittings, and hence the absolute neutrino mass scale is still to be determined. The lifetimes of the neutrinos are also weakly constrained compared to the other Standard Model (SM) particles.

Cosmology stands at present as one of the most powerful probes of neutrino properties. Indeed, one key prediction of the standard $\Lambda$CDM cosmological model is the existence of a cosmic neutrino background (\cnb), formed approximately one second after the Big Bang. These relic neutrinos make a non-negligible contribution to the radiation budget in the early universe and to the matter budget in the late universe, leaving measurable imprints on many cosmological observables. Strong indirect evidence of the \cnb~arises from the constraints on the effective number of relativistic species $N_{\rm eff}$ around the Big Bang nucleosynthesis (BBN) \cite{Pisanti:2020efz, Fields:2019pfx} and cosmic microwave background (CMB) \cite{Follin:2015hya, Baumann:2019keh} eras. Moreover, in recent years the combination of CMB and Large Scale Structure (LSS) observations has yielded increasingly stringent constraints on the  absolute neutrino mass scale. In particular, the latest measurements by the DESI collaboration have led to the tightest upper bound to date on the total neutrino mass, $\sum m_\nu$ < 0.064 eV (95$\%$) \cite{DESI:2025zgx, DESI:2025ejh}, obtained by combining their baryon acoustic oscillation (BAO) data with \Planck~\cite{Planck:2018vyg,Rosenberg:2022sdy, Carron:2022eyg} and ACT \cite{ACT:2023kun,ACT:2023dou} CMB data\footnote{Let us note that this neutrino mass constraint assumes three degenerate neutrinos \cite{DESI:2025zgx}. This limit is relaxed to $\sum m_\nu <0.101 \ \rm{eV}$ and $\sum m_\nu <0.133 \ \rm{eV}$ when imposing the normal and inverted orderings, respectively, and to $\sum m_\nu <0.163 \ \rm{eV}$ when allowing for dynamical dark energy with a $w_0w_a$ equation of state \cite{DESI:2025ejh}.}.  This bound, although cosmological model-dependent, is more than an order of magnitude stronger than the direct laboratory bound of  $\sum m_\nu$ < 1.35 eV (90$\%$) from the KATRIN tritium-beta-decay experiment \cite{KATRIN:2021uub, KATRIN:2024cdt}, and is approaching the lower limit from oscillation experiments, $\sum m_\nu > 0.059 \ \rm{eV}$ \cite{deSalas:2020pgw,Esteban:2024eli}.  \ 

The tightest bounds on neutrino lifetime also come from cosmology. In the case of radiative decays, where the final states can interact electromagnetically, the bounds on CMB spectral distortions have been used to place very strong lower limits on the neutrino lifetime, 
namely $\tau_\nu > (10^2-10^4) \tU$ \cite{Aalberts:2018obr}, where $\tU =13.8 \ \mathrm{Gyr} = 4.35 \times 10^{17}~\mathrm{s}$ denotes the age of the universe. In constrast, the limits on invisible decay channels are much weaker. For neutrinos decaying before recombination (while still ultra-relativistic), decay and inverse decay processes can prevent neutrinos from free-streaming, altering the CMB temperature power spectrum at $\ell \gtrsim 200$ \cite{Taule:2022jrz}. This fact has been exploited to place constraints on neutrino lifetime, $\tau_\nu \gtrsim  4\times 10^6~\mathrm{s}~ (m_\nu/0.05 \mathrm{eV} )^5$, in the case of decays to dark radiation \cite{Barenboim:2020vrr} (see \cite{Basboll:2008fx, Archidiacono:2013dua, Escudero:2019gfk} for earlier work).  If one considers ultra-relativistic decays into lighter neutrinos in a manner consistent with the measured mass splittings, the lifetime bounds are generally weakened to $\tau_\nu \gtrsim (10^2-10^7)~\mathrm{s}$, depending on the decay mass gap and number of decay channels \cite{Chen:2022idm}. On the other hand, non-relativistic neutrinos decaying into dark radiation with lifetimes $\tau_\nu \sim 0.01-0.1 \tU$ have been invoked as a means to relax the cosmological neutrino mass bounds \cite{Serpico:2007pt, Chacko:2019nej, Chacko:2020hmh, Escudero:2020ped, FrancoAbellan:2021hdb}. Let us also note that there are neutrino lifetime bounds from Supernova 1987A \cite{Frieman:1987as}, astrophysical neutrinos measured at IceCube \cite{Baerwald:2012kc, Pagliaroli:2015rca, Bustamante:2016ciw, Denton:2018aml, Abdullahi:2020rge, Bustamante:2020niz}, solar neutrinos \cite{Joshipura:2002fb, Beacom:2002cb, Bandyopadhyay:2002qg, Berryman:2014qha}, and atmospheric neutrinos and long baseline experiments \cite{Gonzalez-Garcia:2008mgl, Gomes:2014yua, Choubey:2018cfz, SNO:2018pvg}. However, these constraints are generally far less stringent than the cosmological ones. \ 
Future observations of the diffuse supernova neutrino background~\cite{Ando:2023fcc} might yield tight constraints on the order of $\tau_\nu /m_\nu \gtrsim 10^{10}$~s/eV~\cite{Ando:2003ie, Fogli:2004gy, Ivanez-Ballesteros:2022szu}.

A direct detection of the \cnb~would mark a major breakthrough for cosmology and particle physics, opening a new window onto the early universe. At the same time, it would serve as an important tool for testing non-standard neutrino interactions, such as invisible decay processes. Crucially, its detection would discard the possibility that neutrinos have fully decayed to dark radiation at some stage during the evolution of the universe, and it could likewise constrain decay scenarios between neutrino mass eigenstates with lifetimes of the order $\tau_\nu \sim \tU$ \cite{Akita:2021hqn}. Despite the significant experimental challenges associated to measuring relic neutrinos, future experiments such as PTOLEMY (Princeton
Tritium Observatory for Light, Early-Universe, Massive-Neutrino Yield) \cite{PTOLEMY:2018jst, PTOLEMY:2019hkd} are aiming to detect the \cnb~through neutrino capture on $\beta$-decaying nuclei, in particular tritium \cite{Cocco:2007za}. If the tritium targets are polarized, it may become possible to extract directional information from the neutrino capture rates and measure the anisotropies on the neutrino sky \cite{Lisanti:2014pqa}. These anisotropies carry both primordial and secondary perturbations (imprinted at late-times by the LSS), so they provide a unique probe of the matter distribution \cite{Tully:2022erg} as well as of fundamental neutrino properties. \ 

In this work, we study the effects of neutrino decays on the anisotropies of the \cnb,\footnote{We note that the authors of Ref. \cite{Akita:2021hqn} had previously studied the effects of invisible neutrino decay on relic neutrino capture on tritium, but their analysis was restricted to the isotropic component.} focusing on a framework where neutrinos decay non-relativistically into lighter neutrinos and a massless scalar, in a manner consistent with the experimentally determined mass splittings. To this end, we perform the first full implementation of late-time decays of the type $\nu_H \rightarrow \nu_l +\phi$ within a linear Einstein-Boltzmann solver, and compute the \cnb~angular power spectrum for various lifetimes and decay channels using the solutions of the neutrino Boltzmann hierarchy. We find that neutrino decay models that are undetectable with current CMB data leave huge imprints on the \cnb~power spectra, and consequently they could be tested in a future polarized tritium target run of the PTOLEMY experiment. Our code will also allow to confront this decay scenario against the latest cosmological data, and to investigate a larger class of models where warm dark matter decays into massless and massive particles. \

This paper is structured as follows. In \autoref{sec:theory_framework} we describe the framework of invisible neutrino decay, and present the Boltzmann equations at the background and linear perturbation level, together with the formalism to compute the \cnb~angular spectrum. In \autoref{sec:numerical_implementation}, we discuss our numerical implementation, the classification of the neutrino decay scenarios and the effects on the background daughter distribution. In \autoref{sec:results}, we report our main results for the CMB and \cnb~temperature anisotropy spectra. We conclude in \autoref{sec:conclusion} and discuss an outlook to future developments. For completeness, in App. \ref{sec:app_a} and App. \ref{sec:app_b} we provide derivations for the background and perturbed momentum-integrated collision terms of the massless particle $\phi$. 

\section{Theoretical framework}
\label{sec:theory_framework}

\subsection{The physical system}

In this work, we consider the 2-body decay of a massive active neutrino $\nu_H$ into a lighter active neutrino $\nu_l$ and a massless scalar particle $\phi$. 
Following Ref.~\cite{Barenboim:2020vrr}, we assume that this non-standard interaction is described by the effective Lagrangian 
\begin{equation}
\pazocal{L}_{\text{int}}=\mathfrak{g}_{i j} \bar{\nu}_i \nu_j \phi.
\label{eq:Lagrangian}
\end{equation}
Here $i$ and $j$ label mass eigenstates, and $\mathfrak{g}_{i j} = \mathfrak{g}$ is a universal coupling constant.  The scalar $\phi$ population is assumed to be produced at late times only through neutrino decay. We concentrate on the regime of small coupling strengths $\mathfrak{g}$, for which $2 \leftrightarrow 2$ scattering processes are always irrelevant and the decay process $\nu_H \rightarrow \nu_l + \phi$ dominates \cite{Barenboim:2020vrr}. The corresponding rest-frame decay rate is given by  \cite{Escudero:2020ped}
\begin{equation}
\Gamma = \frac{\left(\mH-\ml \right)\left(\mH + \ml \right)^3}{4 \pi \mH^3} \mathfrak{g}^2,
\end{equation}
where $\mH$ and $\ml$ are the masses of the heavier and lighter neutrino, respectively, and for concreteness it is assumed that neutrinos are Majorana particles.\footnote{For Dirac particles, the decay rate can simply be obtained as $\Gamma^{\text{Dirac}} = \frac{1}{4} \Gamma^{\text{Majorana}}$.}

Within the small-$\mathfrak{g}$ regime, we focus on `late-time' decays, meaning that neutrinos are assumed to decay after becoming non-relativistic. Hence, the contribution from inverse decay processes is kinematically suppressed and can be safely neglected.
The region of non-relativistic decays is set by the threshold condition on the rest-frame neutrino lifetime $\tau_\nu > H^{-1} (a_\nr)$, where $a_\nr$ is the approximate scale factor at which neutrinos transition to a non-relativistic regime, and is defined as $3T_{\nu}\left( a_\nr \right) = \mH$. The Hubble scale at $a_\nr$ is given by \cite{FrancoAbellan:2021hdb}
\begin{align}
H\left(a_\nr \right) & =H_0 \sqrt{\Omega_m}\left(\frac{\mH}{3T_{\nu 0}}\right)^{3 / 2} \label{eq:Gamma_thres} \\& \simeq 765 \mathrm{~Gyr}^{-1} \left(\frac{H_0}{68 \mathrm{~km} / \mathrm{s} / \mathrm{Mpc}}\right)\left(\frac{\Omega_m}{0.3}\right)^{1 / 2}\left(\frac{3 \mH}{1 \mathrm{eV}}\right)^{3 / 2}\left(\frac{1.5 \times 10^{-4} \mathrm{eV}}{T_{\nu 0}}\right)^{3 / 2} \nonumber,
\end{align}
where $T_{\nu 0}$ is the present-day neutrino temperature. To ensure that even neutrinos in the high-energy tail of the distribution are non-relativistic during the decay, we have chosen values of $\tau_\nu$ well within the bounds set by \autoref{eq:Gamma_thres}.\ 

Previous works have considered the limit of massless daughter neutrinos, $\ml \rightarrow 0$, meaning that neutrinos decay entirely into a \emph{dark radiation} (DR) fluid.  This scenario predicts a late-time transfer of 
energy from the matter to radiation sector, reducing the impact of neutrinos on the expansion rate and structure growth. This in turn allows to significantly relax the very tight cosmological bounds on neutrino masses \cite{Chacko:2019nej, Chacko:2020hmh, Escudero:2020ped, FrancoAbellan:2021hdb}. However, it presents two crucial limitations:
\begin{enumerate}
    \item It is strictly not compatible with the mass splittings determined by oscillation experiments \cite{deSalas:2020pgw}, unless $\mH \lesssim \sqrt{|\Delta m^2_{32}|} \simeq 0.05 \ \rm{eV}$ (where $\Delta m^2_{32}$ denotes the atmospheric mass gap), or $\nu_l$ is taken to be a very light sterile neutrino instead of an active neutrino \cite{Escudero:2020ped}.  
    \item The strong relaxation of neutrino mass bounds is hardly compatible with a potential direct detection of the \cnb. Indeed, to relax mass bounds appreciably, neutrinos need to decay to DR with lifetimes shorter than the age of the universe $\tau_\nu \lesssim 0.1 \tU$, in which case the \cnb~would have decayed away at present \cite{Serpico:2007pt}. Conversely, if neutrinos decay to DR with a lifetime $\tau_\nu \sim \tU$---such that a sizable population of relic neutrinos still remains today---the impact on the \cnb~anisotropies is very small (as we explicitly checked).   
\end{enumerate}
Therefore, in this work we reinstate a finite mass for the lighter daughter neutrino, since such decays can account for the experimentally determined mass splittings and, as we will show, leave strong signatures on the \cnb~temperature anisotropies of the daughter neutrino.\footnote{The late-time effect of $\nu_H \rightarrow \nu_l + \phi$ on CMB and LSS observables has not yet been studied in the literature. However, this scenario is expected to only slightly alleviate the neutrino mass bounds, since the decay products always contain an active massive neutrino, and therefore the impact on the expansion rate and structure growth is not significantly reduced. We will explore this in a forthcoming publication. }

\subsection{Boltzmann equations}

We now introduce the Boltzmann equations to track the cosmological evolution of the $\{\nu_H, \nu_l, \phi\}$ system at both the homogeneous (background) and the perturbed (first-order) level. We follow the same conventions as \cite{Ma:1995ey}. We choose to work in the synchronous gauge, whose line element is given by
\begin{equation}
    ds^2=a(\tau)^2[-d\tau^2+(\delta_{i j}+h_{i j}(\boldsymbol{x}, \tau)) d x^i d x^j],
\label{eq:metric}
\end{equation}
where $a(\tau)$ is the scale factor and $h_{i j}(\boldsymbol{x}, \tau)$ are the scalar perturbations to the metric. We split the phase space distribution (PSD) of the $i$th particle into a homogeneous and isotropic background contribution plus a linear perturbation
\begin{equation}
    f_i(\boldsymbol{x}, \boldsymbol{q}_i, \tau)=\bar{f}_i(q_i, \tau)\left[1+\Psi_i(\boldsymbol{x}, \boldsymbol{q}_i, \tau)\right],
\label{eq:PSD}
\end{equation}
where $\boldsymbol{x}$ are the spatial coordinates, $\boldsymbol{q}_i$ is the comoving momentum (with $q_i \equiv |\boldsymbol{q}_i|$), and $\tau$ denotes conformal time. The evolution of the PSD is dictated by the Boltzmann equation, given in relativistic notation by 
\begin{equation}
    P^{\mu} \frac{\partial f_i}{\partial x^{\mu}} - \Gamma^{\nu}_{\rho\sigma}P^{\rho}P^{\sigma} \frac{\partial f_i}{\partial P^{\nu}} = \frac{\epsilon_i}{a^2} \left( \frac{df_i}{d\tau} \right)_C ,
\label{eq:BE}
\end{equation}
where $P^\mu$ is the 4-momentum, $\epsilon_i \equiv \left(q_i^2+a^2 m_i^2\right)^{1 / 2}$ is the comoving energy of species $i$, and $\Gamma^{\nu}_{\rho\sigma}$ are the Christoffel symbols that capture all gravitational physics. On the r.h.s.\ is the collision term, which was derived at zeroth- and first-order for the neutrino decay and inverse processes $\nu_H \leftrightarrow \nu_l +\phi$ in Ref.~\cite{Barenboim:2020vrr}. In the following, we present these Boltzmann equations in the non-relativistic limit, i.e., excluding inverse decay and quantum statistics terms. 

\subsubsection{Background equations}
\label{sec:back_eqs}
 
The zeroth-order Boltzmann equations for the parent neutrino $\nu_H$, the daughter neutrino $\nu_l$ and the massless scalar $\phi$ are given in Eqs. 4.12 - 4.14 of Ref.~\cite{Barenboim:2020vrr}. In the non-relativistic limit, these equations reduce to 
\begin{align}
\frac{\partial \bar{f}_{\nu H}\left(q_1\right)}{\partial \tau} &=-\frac{a^2 \mH \Gamma}{\epsilon_1} \bar{f}_{\nu H}\left(q_1\right), \label{eq:BE_background_h} \\
\frac{\partial \bar{f}_{\nu l}\left(q_2\right)}{\partial \tau} &=\frac{a^2 \mH^3 \Gamma}{\left(\mH^2-\ml^2\right) \epsilon_2 q_2} \int_{q_{1-}^{(\nu l)}}^{q_{1+}^{(\nu l)}} dq_1 \frac{q_1}{\epsilon_1} \bar{f}_{\nu H} (q_1),
\label{eq:BE_background_l} \\ 
\frac{\partial \bar{f}_{\phi} \left(q_3\right)}{\partial \tau} &=\frac{2 a^2 \mH^3 \Gamma}{\left(\mH^2 - \ml^2\right) q_3^2} \int_{q_{1-}^{\phi}}^{\infty} dq_1 \frac{q_1}{\epsilon_1} \bar{f}_{\nu H} \left(q_1\right),
\label{eq:BE_background_dr}
\end{align}
where the integration limits are given by 
\begin{align}
q_{1 \pm}^{(\nu l)} &=\left|\frac{\epsilon_2\left(\mH^2-\ml^2\right) \pm q_2\left(\mH^2+\ml^2\right)}{2 \ml^2}\right|,
\label{eq:q1_ml}  \\ 
q_{1-}^{(\phi)} &= \left|\frac{a^2\left(\mH^2-\ml^2\right)^2-4 \mH^2 q_3^2}{4 q_3\left(\mH^2-\ml^2\right)}\right|. 
\label{eq:q1_phi}
\end{align}
In past works, the two decay products $\{\nu_l, \phi\}$ were assumed to be massless, and hence could be combined into a single DR fluid (e.g. see \cite{Holm:2022eqq}). The benefit of this approach is that the momentum degrees of freedom of the decay products can be integrated out, and one only needs to track the evolution of the background density and momentum averaged perturbations of the DR, significantly reducing the number of equations to evolve. In our case, the daughter neutrino $\nu_l$ is massive, so the momentum integration can only be performed for the scalar particle $\phi$. To get the evolution of the DR energy density, we thus integrate \autoref{eq:BE_background_dr} over $4\pi a^{-4}dq_3q_3^3$, and follow similar steps as in App. A of Ref.~\cite{Holm:2022eqq}, but using $ \bar{f}_{\dr} = \frac{1}{2}\bar{f}_{\phi}$.\footnote{The factor $1/2$ compensates the fact that we have implicitly removed a spin-factor 2 from $\nu_H$.} Our full derivation is presented in App. \ref{sec:app_a}. In this way, we arrive at 
\begin{equation}
\dot{\rho}_{\mathrm{dr}}  +4 a H \rho_{\mathrm{dr}}= \varepsilon a \Gamma \mH n_{\nu H},
\label{eq:eom_density_dr}
\end{equation}
where dots indicate derivatives with respect to conformal time, $n_{\nu H}$ indicates the number density of the decaying neutrino, and we have defined 
\begin{equation}
\varepsilon \equiv \frac{1}{2} \left(1-\frac{\ml^2}{\mH^2}\right).
\label{eq:epsilon}
\end{equation}
We remark that \autoref{eq:eom_density_dr} is identical to the analogous expression in \cite{Holm:2022eqq}, except for the extra factor $\varepsilon$. This result is physically sensible, as $\varepsilon$ represents the fraction of rest-mass energy of $\nu_H$ transferred to the DR. Indeed, the momentum integration of \autoref{eq:BE_background_h} and \autoref{eq:BE_background_l} yields
\begin{align}
\dot{\rho}_{\nu H} +3 a H\left(\rho_{\nu H}+p_{\nu H}\right) &=-a \Gamma \mH n_{\nu H},
\label{eq:eom_density_h} \\
\dot{\rho}_{\nu l} +3 a H\left(\rho_{\nu l}+p_{\nu l}\right) &= \left(1-\varepsilon\right) a \Gamma \mH n_{\nu H},
\label{eq:eom_density_l}
\end{align}
where $p_{\nu H}$ and $p_{\nu l}$ denote the pressure of the parent and daughter neutrino, respectively. From this, we clearly see that the factor $\varepsilon$ accounts for energy conservation. In the limit $\ml \rightarrow 0$, we get $\varepsilon = \frac{1}{2}$ and $p_{\nu l} = \frac{1}{3}\rho_{\nu l}$, so the energy is equally distributed among the two daughter species (which follow the exact same equations), and we recover the background DR equation of \cite{Holm:2022eqq}. For the numerical implementation, we use \autoref{eq:eom_density_dr} to directly evolve $\rho_{\dr}$, while for $\{\nu_H,\nu_l\}$ we first track the evolution of  $\bar{f}_{\nu H}$ and $\bar{f}_{\nu l}$ at each momentum bin using \autoref{eq:BE_background_h}-\autoref{eq:BE_background_l}, and then integrate over momenta to get $\rho_{\nu H}$ and $\rho_{\nu l}$.\footnote{Note that $\rho_{\nu H}$ and $\rho_{\nu l}$ cannot be obtained by directly evolving \autoref{eq:eom_density_h} and \autoref{eq:eom_density_l}, since $p_{\nu H}$, $n_{\nu H}$ and $p_{\nu l}$ require the knowledge of $\bar{f}_{\nu H}$ and $\bar{f}_{\nu l}$ at each momentum bin. As we will see later, this is also required to trace the evolution of perturbations for the neutrino species.} Regarding the initial conditions, we take a relativistic Fermi-Dirac distribution as predicted by standard neutrino decoupling 
\begin{equation}
 \bar{f}_{\nu H} (q, \tau_i ) =    \bar{f}_{\nu l} (q, \tau_i) = \frac{1}{e^{q/T_{\nu 0}}+1}. 
\end{equation}

\subsubsection{Perturbation equations}
\label{sec:pert_eqs}

Following the usual procedure, we decompose the Fourier transform of the first-order perturbed part of the PSD, $\Psi_i (k,q_i,\hat{k} \cdot \hat{q}_i, \tau )$, in terms of Legendre polynomials 
\begin{align}
    \Psi_i(k, q_i, \hat{k} \cdot \hat{q}_i, \tau) &= \sum_{\ell=0}^{\infty}(-i)^{\ell}(2 \ell+1) \Psi_{i, \ell}\left(k, q_i, \tau \right) P_{\ell}(\hat{k} \cdot \hat{q}_i).  \label{eq:Leg_expansion} 
\end{align}
By combining \autoref{eq:Leg_expansion} with the first-order Boltzmann equation for $\Psi_i$ (see Eq. 4.18 in \cite{Barenboim:2020vrr}), we obtain an infinite hierarchy of equations of motion for the multipole moments $\Psi_{i, \ell}(k, q_i,\tau)$, 
\begin{align}
\dot{\Psi}_{i, 0}\left(q_i\right) & =-\frac{q_i k}{\epsilon_i} \Psi_{i, 1}\left(q_i\right)+\frac{\dot{h}}{6} \frac{\partial \ln \bar{f_i}}{\partial \ln q_i} +\pazocal{C}_0^{(1)}\left[\Psi_i\left(q_i\right)\right], \nonumber \\
\dot{\Psi}_{i, 1}\left(q_i\right) & =\frac{q_i k}{\epsilon_i}\left(-\frac{2}{3} \Psi_{i, 2}\left(q_i\right)+\frac{1}{3} \Psi_{i, 0}\left(q_i\right)\right)+\pazocal{C}_1^{(1)}\left[\Psi_i\left(q_i\right)\right], \label{eq:BH} \\
\dot{\Psi}_{i, 2}\left(q_i\right) & =\frac{q_i k}{\epsilon_i}\left(-\frac{3}{5} \Psi_{i, 3}\left(q_i\right)+\frac{2}{5} \Psi_{i, 1}\left(q_i\right)\right)-\frac{\partial \ln \bar{f}_i}{\partial \ln q_i}\left(\frac{2}{5} \dot{\eta}+\frac{1}{15} \dot{h}\right)+\pazocal{C}_2^{(1)}\left[\Psi_i\left(q_i\right)\right], \nonumber \\
\dot{\Psi}_{i, \ell>2}\left(q_i\right) & =\frac{k}{2 \ell+1} \frac{q_i}{\epsilon_i}\left[\ell \Psi_{i, \ell-1}\left(q_i\right)-(\ell+1) \Psi_{i, \ell+1}\left(q_i\right)\right]+\pazocal{C}_{\ell}^{(1)}\left[\Psi_i\left(q_i\right)\right], \nonumber
\end{align}
where the effective first-order collision term,
\begin{equation}
\pazocal{C}_{\ell}^{(1)}\left[\Psi_i\left(q_i\right)\right] \equiv \frac{1}{\bar{f_i}}\left(\frac{df_i}{d\tau}\right)_{C, \ell}^{(1)}-
\frac{\dot{\bar{f}}_i}{\bar{f_i}} \Psi_{i, \ell}, 
\label{eq:coll_term}
\end{equation}
is composed of the $\ell$ order Legendre moment of the first-order collision term and the background evolution term. The Boltzmann hierarchy in \autoref{eq:BH} is in the same form as the one for standard free-streaming neutrinos \cite{Ma:1995ey}, except for the collision terms and the fact that the partial derivative $\partial \ln \bar{f}_i /\partial\ln q_i$ is now time-dependent. For the decaying neutrino $\nu_H$, the two components in \autoref{eq:coll_term} exactly cancel \cite{Barenboim:2020vrr}, leaving no collision term when excluding inverse decays. This reflects the fact that the decay is a pure background process. For the daughter particles $\nu_l$ and $\phi$, the collision terms are given in Eqs. 4.25 - 4.26 of Ref. \cite{Barenboim:2020vrr}. In the non-relativistic limit, these reduce to 
\begin{align}
\mathcal{C}_{\ell}^{(1)}[\Psi_{\nu l}(q_2)] &= \frac{a^2 \mH^3 \Gamma}{(\mH^2-\ml^2)\epsilon_2 q_2 \bar{f}_{\nu l}(q_2)} \int_{q_{1-}^{(\nu l)}}^{q_{1+}^{(\nu l)}} dq_1 \frac{q_1}{\epsilon_1} \bar{f}_{\nu H} (q_1) ( \Psi_{\nu H,\ell}(q_1)P_{\ell}(\cos \alpha^*)-\Psi_{\nu l,\ell}(q_2)), \label{eq:C_psi_nul} \\
\pazocal{C}_{\ell}^{(1)}\left[\Psi_{\phi}\left(q_3\right)\right] &= \frac{2a^2 \mH^3 \Gamma}{\left(\mH^2 - \ml^2 \right)q_3^2 \bar{f}_{\phi}\left(q_3\right)} \int_{q_{1-}^{(\phi)}}^{\infty} dq_1 \frac{q_1}{\epsilon_1} \bar{f}_{\nu H}\left(q_1\right)\left(\Psi_{\nu H, \ell}\left(q_1\right) P_{\ell}\left(\cos \beta^*\right)-\Psi_{\phi, \ell}\left(q_3\right)\right), \label{eq:C_psi_phi}
\end{align}
with
\begin{align}
\cos \alpha^* &=\frac{2 \epsilon_1 \epsilon_2-a^2\left(\mH^2+\ml^2\right)}{2 q_1 q_2}, \label{eq:cos_alpha} \\
\cos \beta^* &=\frac{2 \epsilon_1 q_3-a^2\left(\mH^2-\ml^2\right)}{2 q_1 q_3}. \label{eq:cos_beta}
\end{align}
Similarly to what we did at the background level, we can 
eliminate the momentum degrees of freedom of the massless scalar $\phi$ by taking the momentum averaged DR perturbation,
\begin{equation}
    F_{\mathrm{dr}}(k, \tau) \equiv r_{\mathrm{dr}} \frac{\int dq_3  q_3^3 ~\bar{f}_{\phi}(q_3, \tau) \Psi_{\phi}(k, q_3,\hat{k} \cdot \hat{q}_3, \tau)}{\int dq_3 q_3^3 ~\bar{f}_{\phi}(q_3, \tau)},
\label{eq:integrated_pert}
\end{equation}
where we have adopted the convention $r_{\mathrm{dr}} \equiv \rho_{\mathrm{dr}} a^4 / \rho_{\text{crit},0}$, as in previous works \cite{FrancoAbellan:2021sxk, Holm:2022eqq}. After decomposing \autoref{eq:integrated_pert} in Legendre polynomials, and combining with the Boltzmann hierarchy in \autoref{eq:BH}, the resulting infinite hierarchy of equations for the multipole moments $F_{\mathrm{dr}, \ell}$ reads 
\begin{align}
    & \dot{F}_{\mathrm{dr}, 0}=-k F_{\mathrm{dr}, 1}-\frac{2}{3} r_{\mathrm{dr}} \dot{h}+\left(\frac{dF_{\mathrm{dr}}}{d\tau}\right)_{C, 0}^{(1)}, \nonumber\\
    & \dot{F}_{\mathrm{dr}, 1}=\frac{k}{3} F_{\mathrm{dr}, 0}-\frac{2 k}{3} F_{\mathrm{dr}, 2}+\left(\frac{dF_{\mathrm{dr}}}{d\tau}\right)_{C, 1}^{(1)}, \label{eq:integrated_BH} \\
    & \dot{F}_{\mathrm{dr}, 2}=\frac{2 k}{5} F_{\mathrm{dr}, 1}-\frac{3 k}{5} F_{\mathrm{dr}, 3}+\frac{4}{15} r_{\mathrm{dr}}(\dot{h}+6 \dot{\eta})+\left(\frac{dF_{\mathrm{dr}}}{d\tau}\right)_{C, 2}^{(1)}, \nonumber \\
    & \dot{F}_{\mathrm{dr}, \ell \geq 3}=\frac{k}{2 \ell+1}\left(\ell F_{\mathrm{dr}, \ell-1}-(\ell+1) F_{\mathrm{dr}, \ell+1}\right)+\left(\frac{dF_{\mathrm{dr}}}{d\tau}\right)_{C, \ell}^{(1)}. \nonumber
\end{align}
To obtain the DR first-order collision term $(dF_{\mathrm{dr}}/ d\tau)_{C, \ell}^{(1)}$ from \autoref{eq:C_psi_phi}, we just reproduce the calculations performed in App. C of Ref.~\cite{Holm:2022eqq}, but using $\Psi_{\dr} = \Psi_{\phi}$. Our complete derivation is presented in App. \ref{sec:app_b}. We arrive at the expression
\begin{equation}
    \left(\frac{dF_{\mathrm{dr}}}{d\tau}\right)_{C, \ell}^{(1)}=\dot{r}_{\mathrm{dr}} \frac{\int_0^{\infty} dq_1 q_1^2 \bar{f}_{\nu H}(q_1) \Psi_{\nu H, \ell}(q_1) \pazocal{F}_{\ell}\left(\frac{q_1}{\epsilon_1}\right)}{\int_0^{\infty} dq_1 q_1^2 \bar{f}_{\nu H}(q_1)}, 
    \label{eq:integrated_collision}
\end{equation}
where 
\begin{equation}
\pazocal{F}_{\ell}\left(x\right) \equiv \int_{-1}^{+1} \frac{P_\ell (u)}{(1-xu)^3}du, 
\end{equation}
and $\dot{r}_{\mathrm{dr}}$ is given by 
\begin{equation}
    \dot{r}_{\mathrm{dr}} = \varepsilon \frac{r_{\mathrm{dr}} a \Gamma \mH n_{\nu H}}{\rho_{\mathrm{dr}}}.
\label{eq:r_dot_dr}
\end{equation}
The collision term in \autoref{eq:integrated_collision} is identical to the one found in \cite{Holm:2022eqq}, except for the extra factor of $\varepsilon$, which is again accounting for the repartition of energy between the two decay products.

\subsection{\cnb~anisotropies}

To model the \cnb~temperature anisotropies, we use the formalism presented in \cite{Tully:2021key}, which provides a way to compute the \cnb~angular power spectrum given the solutions of the neutrino Boltzmann hierarchy in linear theory. In this work, we consider only the anisotropies of the daughter neutrino $i = \nu_l$, since the parent neutrino $\nu_H$ has fully decayed at present for sufficiently short lifetimes. For notational simplicity, we thus omit the index $i$ in what follows.\ 

The starting point is to parameterize the perturbed PSD $f (\boldsymbol{x}, \boldsymbol{q},\tau)$ in terms of a variable $\Delta$, such that $f =(e^{\frac{q}{T_{\nu 0}(1+\Delta)}} +1)^{-1}$. By expanding the PSD to linear order around a Fermi-Dirac distribution, and comparing to \autoref{eq:PSD}, one arrives at the following relation between $\Delta$ and $\Psi$
\begin{equation}
\Delta=-\left(\frac{d \ln \bar{f}}{d \ln q}\right)^{-1} \Psi. 
\label{eq:Delta_Psi}    
\end{equation}
We emphasize that, despite its resemblance to a temperature perturbation, $\Delta$ should strictly not be interpreted as one. Indeed, from the previous relation, one sees that $\Delta$ depends not only on spacetime but also on momenta, $\Delta = \Delta (\boldsymbol{x}, \boldsymbol{q},\tau)$, which leads to departures from a thermal distribution at the perturbed level (this is generally true for massive neutrinos \cite{Lesgourgues:2013sjj}). Furthermore, to derive \autoref{eq:Delta_Psi}, it is assumed that the background PSD $\bar{f}$ follows a Fermi-Dirac distribution at all times, which is a valid assumption for free-streaming neutrinos, but not for neutrinos participating in the decay. For this reason, we regard $\Delta$ merely as a convenient variable that can be defined through \autoref{eq:Delta_Psi}, which allows us to make connection with previous studies of \cnb~anisotropies \cite{Hannestad:2009xu,Tully:2021key}. \ 

After decomposing the Fourier transforms of $\Delta$ and $\Psi$ into Legendre polynomials, one can use \autoref{eq:Delta_Psi} to relate the temperature multipoles $\Delta_\ell (k, q, \tau)$ to the solutions of the neutrino Boltzmann hierarchy, $\Psi_\ell (k, q, \tau)$.\footnote{Ref.~\cite{Tully:2021key} developed a line-of-sight approximation that allows to compute $\Delta_{\ell}(k, q, \tau_0)$ up to a very large multipole $\ell_{\rm max}$ without having to solve a Boltzmann hierarchy of $(\ell_{\rm max}+1)$ equations. While this alternative method is considerably faster, it cannot be easily extended to the decaying neutrino case, and doing so is beyond the scope of our work.} With this, the angular power spectrum for a given neutrino comoving momentum $q$ can be obtained as follows \cite{Tully:2021key}
\begin{equation}
    C_{\ell} (q) = 4 \pi A_s T_{\nu 0}^2 \int d \ln k \left(\frac{k}{k_{\star}}\right)^{n_s-1}\Delta_{\ell}^2(k, q, \tau_0),
\label{eq:power_spectrum_q}
\end{equation}
where $A_s$ is the amplitude of the primordial spectrum, $n_s$ is the spectral index, and the reference wavenumber is fixed to $k_{\star} = 0.05 \ \rm{Mpc}^{-1}$. We perform the integral in \autoref{eq:power_spectrum_q}  for a range of wavenumbers $ 10^{-4} \leq k/\rm{Mpc}^{-1} \leq 10^{-1}$, which are hence in the linear regime of present-day perturbations. The power spectra are computed up to $\ell_{\text{max}} = 17$, which is the multipole at which we truncate the neutrino Boltzmann hierarchy. This $\ell_{\text{max}}$ value is chosen due to computational limitations, but we note that multipoles $\ell \leq  15$ are also the ones that receive contributions mainly from linear scales for neutrino masses $m_\nu \leq 0.05 \ \mathrm{eV}$ \cite{Tully:2021key}, making our approach self-consistent.  Moreover, the initially poor angular resolution of a future directional \cnb~experiment like PTOLEMY implies that it will be sensitive only to large-scale variations.\

The rates for relic neutrino capture are not strongly sensitive to the incoming momentum of the captured neutrinos \cite{Cocco:2007za}.
Therefore, we eliminate the $q$-dependence from $C_\ell (q)$ using the momentum averaging introduced in Ref.~\cite{Hannestad:2009xu}
\begin{equation}
    C_{\ell} = \left[\frac{\int dq~ q^2 \epsilon(q) \bar{f}_{\nu l}(q) \sqrt{C_{\ell}(q)}}{\int dq~ q^2 \epsilon(q) \bar{f}_{\nu l}(q)} \right]^2.
\label{eq:power_spectrum_avg}
\end{equation}
Other choices of average angular power spectra are of course possible, and the appropriate definition ultimately depends on how measurements are performed in a neutrino capture experiment. For instance, Ref.~\cite{Tully:2021key} proposes a similar $q$-bin averaging, but the $q$-dependence is integrated out directly at the level of the temperature multipoles $\Delta_\ell$. Here, we have adopted the $q$-averaging of \autoref{eq:power_spectrum_avg} as it allows for a more straightforward interpretation of the results. Let us also note that the average involves the present-day background PSD of the daughter neutrino $\bar{f}_{\nu l} (q)$, which  generally departs from a Fermi-Dirac distribution due to the neutrino decay process (see \autoref{sec:numerical_implementation}). This directly impacts the weights given to the different $q$-dependent angular power spectra $C_{\ell} (q)$ in \autoref{eq:power_spectrum_avg}, as well as the maximum momentum $q_{\rm max}$ up to which we carry out the momentum integrals.

\section{Numerical implementation}
\label{sec:numerical_implementation}

We have implemented the equations described in \autoref{sec:back_eqs} and \autoref{sec:pert_eqs} in our modified version\footnote{\href{https://github.com/GuillermoFrancoAbellan/CLASSpp_nuDecay}{\texttt{https://github.com/GuillermoFrancoAbellan/CLASSpp\_nuDecay}}.} of the public Einstein-Boltzmann solver \texttt{CLASS} \cite{Lesgourgues:2011re, Blas:2011rf}. In particular, we use a modification of \texttt{CLASS++}, a version of \texttt{CLASS} written in \texttt{C++} that was developed by the authors of Ref.~\cite{Holm:2022eqq} to model the effects of warm dark matter (WDM) decaying non-relativistically into DR.\ 

In the original \texttt{CLASS++} version, one can pass the parameter \texttt{N\_ncdm\_decay\_dr}, which specifies the number of non-cold dark matter (NCDM) species (e.g. WDM, neutrinos) decaying into DR. Each of these species is described by different quantities that can be passed as input parameters, including an initial abundance, a mass, a decay rate, as well as various precision settings. To incorporate the massive daughter, we adopted the following strategy: we set $\texttt{N\_ncdm\_decay\_dr}=2$, and introduced a flag ($\texttt{has\_ncdm\_decay\_dr\_ncdm}$) that specifies whether the decay products include a massive component. When this flag is activated, the code interprets the first NCDM species as the parent particle, and the second NCDM species as the daughter particle, whose equations now include the collision integrals in \autoref{eq:BE_background_l} and \autoref{eq:C_psi_nul}.  The code further imposes that only a single DR species is present (rather than two), with its equations modified to include the $\varepsilon$ factor, as in \autoref{eq:eom_density_dr} and \autoref{eq:r_dot_dr}. In order to close the Friedmann equation, both the initial abundance of the decaying species (\texttt{Omega\_ini\_dncdm}) and the final abundance of the three species (\texttt{Omega\_dncdm\_dr\_ncdm}) must be known.  Our code determines one from the other in a self-consistent manner using a shooting algorithm.\ 

\begin{table}[h!]
\centering
\renewcommand{\arraystretch}{1.3}
\begin{tabular}{|c|c|c|c|c|c|}
\hline
 & Mass ordering & Free-streaming & Decay channel & Decay mass gap \\
\hline
\multicolumn{5}{|c|}{\textbf{Scenario A: one decay channel}} \\
\hline
A2 & NO & $\nu_3$ & $ \nu_2 \rightarrow \nu_1$ & $\Delta m^2_{21}$ \\ \hline
A3 & IO & $\nu_3$ & $\nu_2 \rightarrow \nu_1$ & $\Delta m^2_{21}$ \\
\hline
\multicolumn{5}{|c|}{\textbf{Scenario B: two decay channels}} \\
\hline
B1 & NO & $-$ & $\nu_3 \rightarrow \nu_2, \nu_1$ & $|\Delta m^2_{32}|$ \\ \hline
B2 & IO & $-$ & $\nu_1, \nu_2 \rightarrow \nu_3$ & $|\Delta m^2_{32}| $ \\
\hline
\end{tabular}
\caption{All decay scenarios considered in this work. These are classified according to the number of decay channels (single-decay channel or two-decay channel with the degenerate-$\nu_{1,2}$ approximation), the mass ordering (normal or inverted), and the decay mass gap (solar or atmospheric). Scenarios A2 and A3 involve one free-streaming neutrino. The nomenclature of these scenarios is borrowed from Ref.~\cite{Chen:2022idm}. }
\label{tab:decay_scenarios}
\end{table}

\subsection{Neutrino decay scenarios}

While our \texttt{CLASS++} version in principle allows to test any dark sector model predicting a late-time decay of warm particles into massive and massless components, we optimize the code for the decaying neutrino model described by the Lagrangian in \autoref{eq:Lagrangian}. In practice, we fix the initial NCDM abundances to ensure that the effective number of relativistic neutrino species in the early Universe is always the standard value $N_{\rm eff}^{\rm SM} = 3.044$ \cite{Bennett:2020zkv}, and we introduce several input parameters to enforce consistency of neutrino masses $m_i$ with oscillation data \cite{deSalas:2020pgw}. In particular, the user can specify a mass ordering, i.e., normal ordering (NO; $m_3 > m_2 > m_1 $) or inverted ordering (IO; $m_2 > m_1 > m_3 $), together with the lightest neutrino mass ($m_1$ in NO and $m_3$ in IO), and the code automatically determines the masses of the remaining two neutrino species. Additionally, the user can specify whether the mass gap between parent and daughter particle is the atmospheric ($|\Delta m^2_{32}| \simeq 2.5 \times 10^{-3} \ \rm{eV}^2 $) or the solar one ($\Delta m^2_{21} \simeq 7.5 \times 10^{-5} \ \rm{eV}^2 $). \

In this regard, we note that the decay scenario presented in \autoref{sec:theory_framework} concerns only two of the three active neutrino mass eigenstates. Unless a specific model provides reasons to exclude one mass eigenstate from the Lagrangian in \autoref{eq:Lagrangian}, we generally expect nonzero decay rates $\Gamma_{i\rightarrow j}$ between all possible pairs for the process $\nu_i \rightarrow \nu_j + \phi$. Nevertheless, from the measured mass splittings one can deduce that the decay rates $\Gamma_{i\rightarrow j}$ follow the patterns
\begin{equation}
 \Gamma_{3\rightarrow 2} \simeq \Gamma_{3\rightarrow 1} \gg \Gamma_{2\rightarrow 1} \, \, \, \,  \rm{(NO)} \, \, \, \, \, \, \mathrm{and} \, \, \, \, \, \, 
\Gamma_{2\rightarrow 3} \simeq \Gamma_{1\rightarrow 3} \gg \Gamma_{2\rightarrow 1} \, \, \, \, \rm{(IO)}, 
\label{eq:deg_condition}
\end{equation}
meaning that $\nu_1$ and $\nu_2$ can be treated as degenerate species, and thus the three-state system can be effectively reduced to a two-state one. For NO mass spectra with $m_1 \gtrsim 0.03 \ \rm{eV}$ and all IO mass spectra, the conditions in \autoref{eq:deg_condition} are satisfied at better than $10\%$ \cite{Barenboim:2020vrr}. In our \texttt{CLASS++} version, we give the option to apply the two-state approximation using the flag $\texttt{is\_ncdm\_decay\_degenerate}$.  When this flag is activated for the normal (inverted) hierarchy, the code multiplies by $2$ all collision terms in the $\nu_H$ ($\nu_l$) and $\phi$ Boltzmann equations, as well as all the momentum-integrated quantities of $\nu_l$ ($\nu_H$).\ 

In \autoref{tab:decay_scenarios} we summarize all the decay scenarios that we consider in this work, adopting the nomenclature of Ref.~\cite{Chen:2022idm}. Specifically, we consider: i) scenarios A2 (NO) and A3 (IO), for a single-decay channel with the solar mass gap; and ii) scenarios B1 (NO) and B2 (IO), for two-decay channels with the atmospheric mass gap using the two-state approximation.\footnote{Ref.~\cite{Chen:2022idm} also considered a scenario dubbed A1, corresponding to a single-decay-channel with the atmospheric mass gap. We omit this scenario for simplicity, since it is somewhat similar to scenarios B1 and B2.}  Let us note that in scenarios A2 and A3, one neutrino does not participate in the decay interaction, and is therefore modeled as a standard free-streaming NCDM species in our \texttt{CLASS++} version.

\begin{figure}[h!]
\centering
\includegraphics[width=0.485\linewidth]{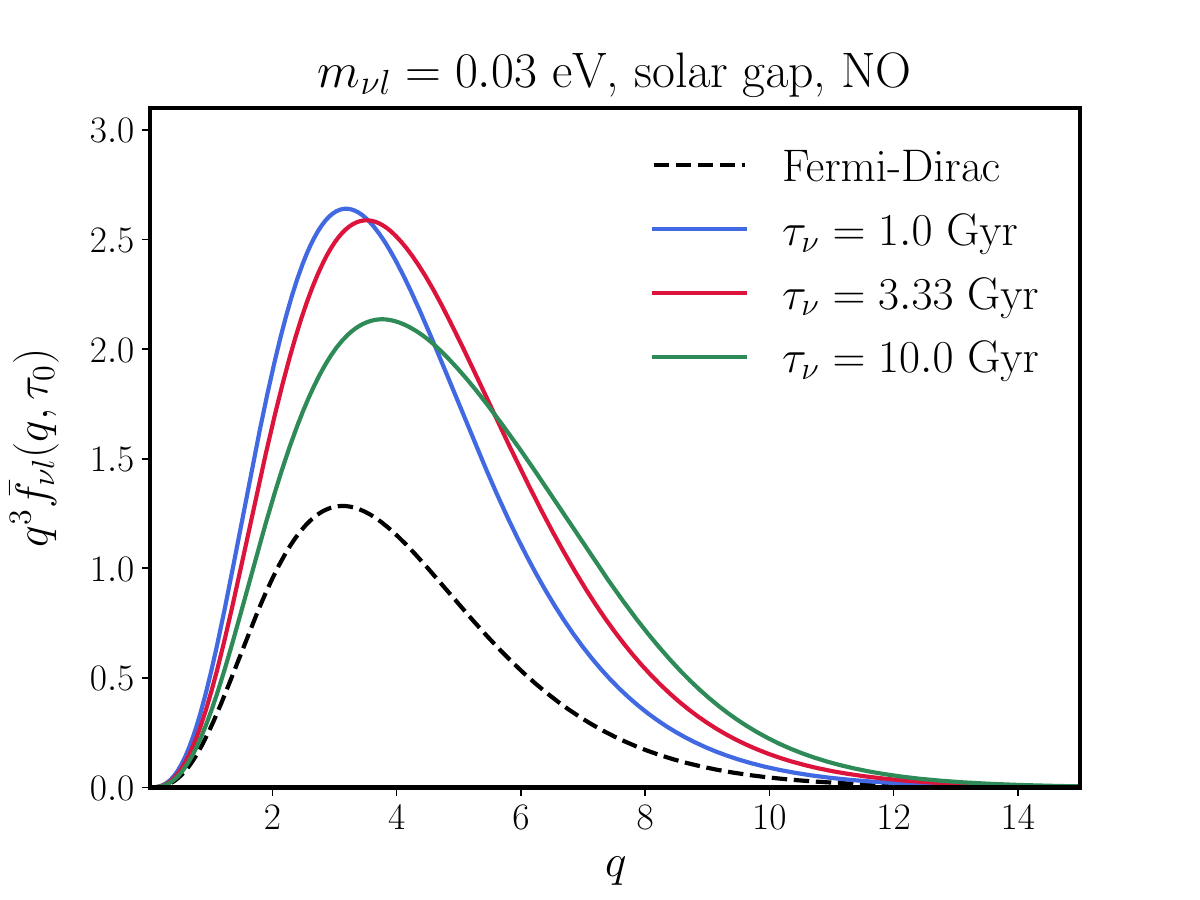} 
\includegraphics[width=0.485\linewidth]{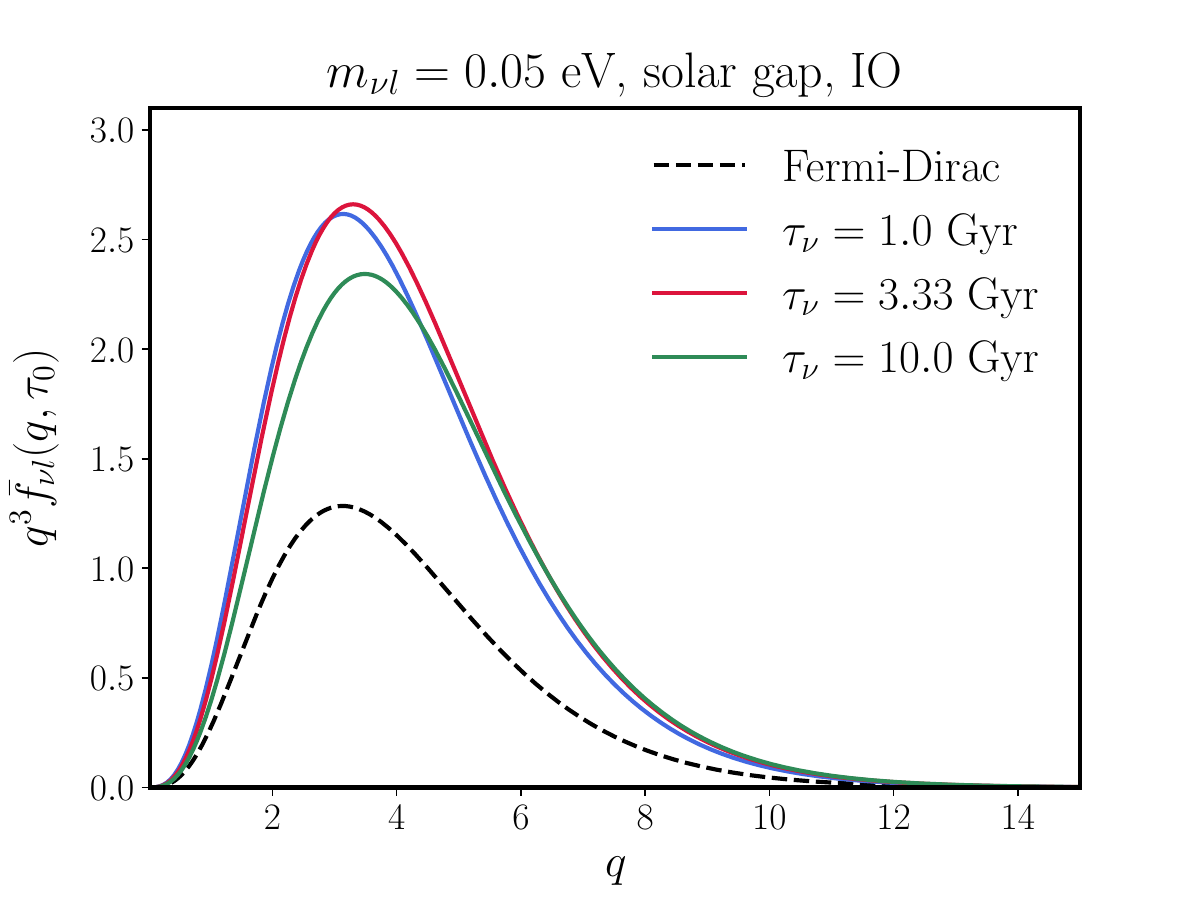} 
\caption{ \small Final phase-space distributions of $\nu_l$ in the neutrino decay scenarios A2 (\textit{left panel}) and A3 (\textit{right panel}) for a range of neutrino lifetimes $\tau_\nu$ and fixed daughter mass $m_{\nu l}$. These are compared to a standard Fermi-Dirac distribution. The comoving momentum $q$ is given in units of $T_{\nu 0}$.
}
\label{fig:PSD_A1_A2}
\end{figure}

\begin{figure}[h!]
\centering
\includegraphics[width=0.485\linewidth]{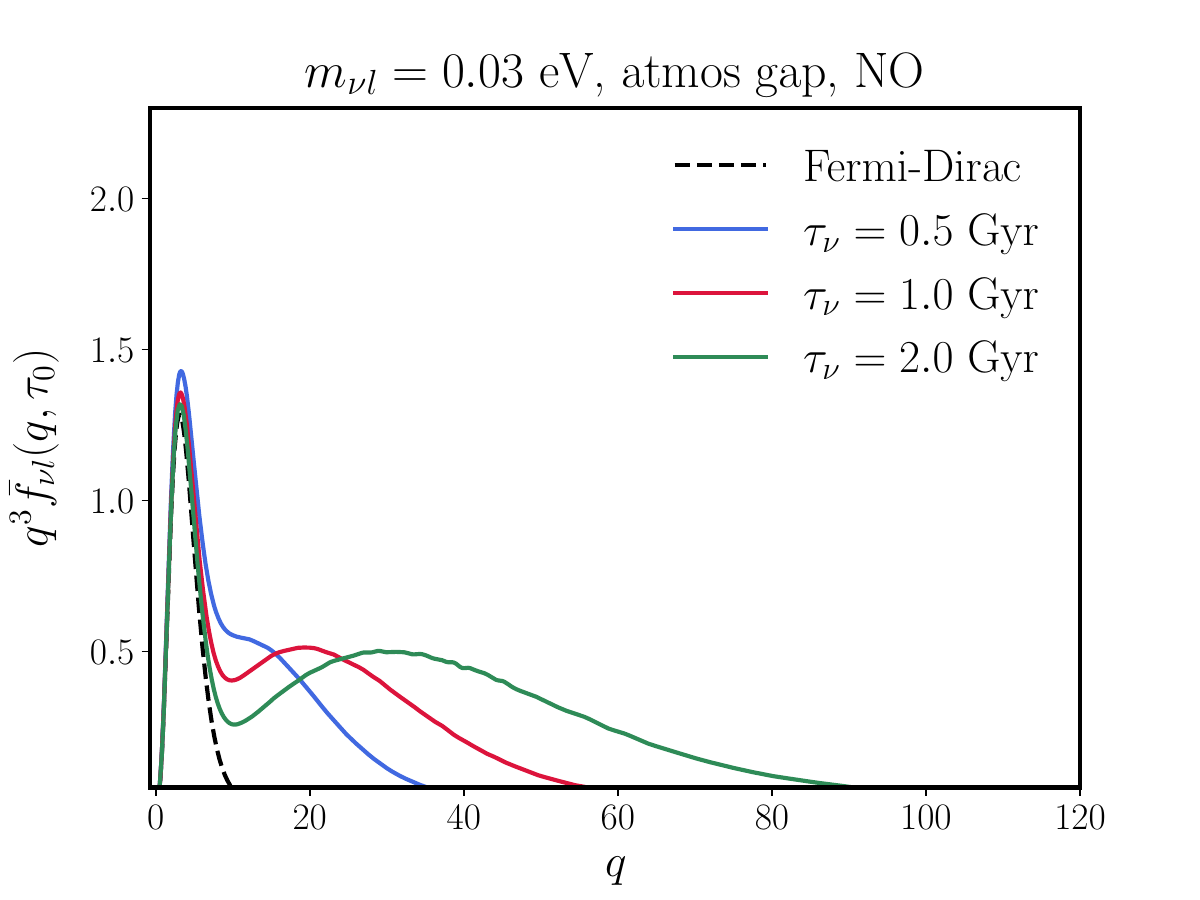} 
\includegraphics[width=0.485\linewidth]{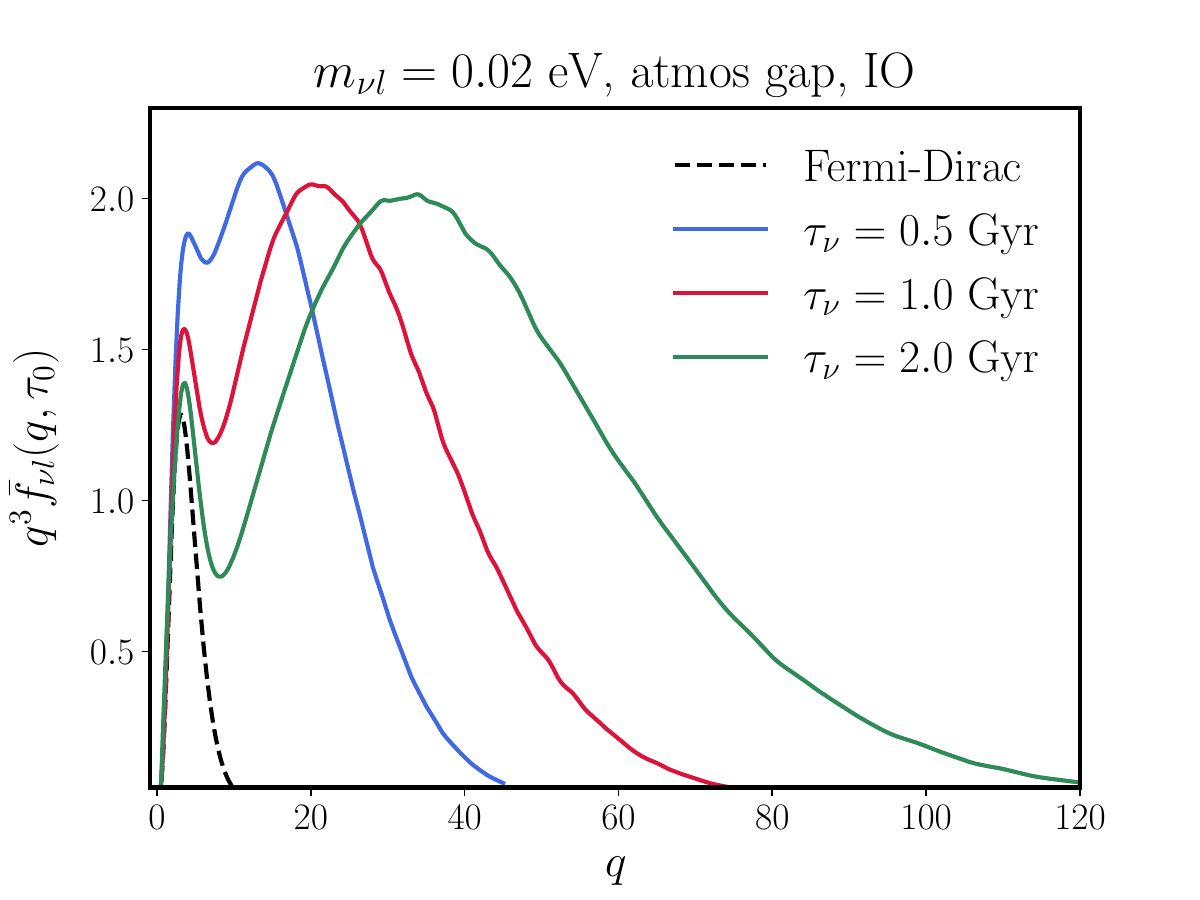} 
\caption{ \small Similar as \autoref{fig:PSD_A1_A2}, but for scenarios B1 (\textit{left panel}) and B2 (\textit{right panel}).}
\label{fig:PSD_B1_B2}
\end{figure} 

\subsection{Background daughter distribution}

Before turning to the discussion of \cnb~anisotropies, it is very instructive to examine how neutrino decay affects the background daughter distribution $\bar{f}_{\nu l}$, since these effects are expected to propagate at the perturbation level. In \autoref{fig:PSD_A1_A2} and \autoref{fig:PSD_B1_B2}, we show the final PSD of $\nu_l$ for all scenarios A and B listed in \autoref{tab:decay_scenarios}, assuming a fixed daughter neutrino mass $\ml$ and several values of neutrino lifetime $\tau_\nu$,\footnote{In the case of B1, the lifetime is computed as $\tau_\nu = 1/(2\Gamma)$, to account for its two distinct decay modes.} and compare with a standard Fermi-Dirac distribution. In each case, the value of $\ml$ is chosen such that the total neutrino mass $\sum m_{\nu}$ remains marginally compatible with the latest \Planck+DESI upper bounds \cite{DESI:2025zgx, DESI:2025ejh}, while also respecting the lower bounds from oscillation data (i.e., $\ml \geq \small{\sqrt{|\Delta m^2_{32}|}}\simeq 0.05 \ \rm{eV}$ for scenario A3) and the limit of validity of the two-state approximation (i.e., $\ml \geq  0.03~\rm{eV}$ for scenario B1).\ 

From \autoref{fig:PSD_A1_A2}-\autoref{fig:PSD_B1_B2}, we see that the present-day $\bar{f}_{\nu l}$ is significantly distorted relative to a Fermi-Dirac distribution, with the strength of the distortion being controlled by $\tau_\nu$. In addition, the final shape of the PSD is heavily influenced by the decay mass gap: in scenarios A (solar gap), the maximum momentum at which $q^3\bar{f}_{\nu l}$ has support coincides with the Fermi-Dirac case ($q_{\rm max}/T_{\nu 0} \sim 15$), whereas in scenarios B (atmospheric gap), the PSD extends to much larger momenta, with longer lifetimes producing longer high-energy tails. This can be understood from the fact that, in the rest-frame of the parent neutrino $\nu_H$, the daughter neutrino $\nu_l$ is emitted with a comoving momentum proportional to the scale factor and the decay mass gap, $q'=a(\mH^2-\ml^2)/(2\mH)$ \cite{Barenboim:2020vrr}. We also see that in the case of B2, $\bar{f}_{\nu l}$ exhibits a stronger amplification relative to a Fermi-Dirac distribution than in B1, because the $\nu_l$ distribution is populated by two decaying $\nu_H$ particles, rather than just one. \ 

Finally, we note that the momentum resolution needed to adequately compute the energy or number densities is also very dependent on the decay mass gap, with scenarios A necessitating a much smaller step size $dq$ than scenarios B. Since the required number of momentum bins $N_q$ and maximum momentum $q_{\rm max}$ vary significantly across the decay parameter space, our \texttt{CLASS++} code includes a function that automatically estimates the optimal values of ($N_q, q_{\rm max}$) for a given $\ml$, $\Gamma$ and neutrino decay scenario. This function can be called by activating the flag \texttt{adjust\_q\_binning}, in which case the values of  $(N_q, q_{\rm max})$ specified in the input file are overwritten by the optimally determined ones. 

\section{Results}
\label{sec:results}

In this section, we study the impact of invisible neutrino decays of the type $\nu_i \rightarrow \nu_j +\phi$ on the CMB and \cnb~anisotropies, and compare the sensitivities of these two probes to such non-standard neutrino interactions. To this end, we use our modified \texttt{CLASS++} version (described in  \autoref{sec:numerical_implementation}) to compute the lensed CMB temperature (TT) power spectrum as well as the \cnb~power spectrum (via \autoref{eq:power_spectrum_avg}) for various neutrino decay models. To isolate the signatures of the decay, we always present residual differences in these power spectra with respect to the equivalent stable scenario, i.e., the one with the same neutrino mass spectrum and $\Gamma = 0$. All other cosmological parameters are fixed to $\{H_0 = 67.37~\mathrm{km}/\mathrm{s}/\mathrm{Mpc},~\omega_b = 0.02233, ~\omega_{\mathrm{cdm}} = 0.1198,~n_s = 0.9652, ~\ln(10^{10}A_s)=3.043,~\tau_{\mathrm{reio}} = 0.0540\}$, which correspond to the \lcdm~best-fit parameters from \Planck~2018 \cite{Planck:2018vyg}.\ 

In order to assess which neutrino decay models may be measurable by the CMB or \cnb, we also show the fractional uncertainties associated to each probe. For the CMB, we take the measurement errors from \Planck~2018. For the \cnb, we assume that the dominant contributions arise from cosmic variance and the PTOLEMY counting statistics for the total number of neutrino capture events.  The 1$\sigma$ uncertainty on $C_\ell$ from cosmic variance is given by \cite{Peebles:1980yev}  
\begin{equation}
\frac{\Delta C_{\ell}}{C_{\ell}} = \sqrt{\frac{2}{2\ell +1}},
\end{equation}
and is larger than 20$\%$ for multipoles $\ell \leq 17$. To compute the fractional uncertainty on $C_\ell$ from the PTOLEMY counting statistics, we follow the same procedure as Ref. \cite{Tully:2022erg}. Namely, we generate a temperature map for the stable limit of each decay scenario, and interpret it as a distribution of neutrino capture events, where each pixel records the number of captures for a total of $N$ neutrino events. The expected count per pixel is then smeared using Poisson statistics, and the map is inverted to find the corresponding power spectrum $C_{\ell}$. This procedure is repeated for a number of trials. We verified that the resulting fractional uncertainty per $\ell$ mode scales as $1/\sqrt{N}$, but in our plots we take $N=10^5$ for concreteness. \ 

In general, we expect the  \cnb~spectrum to be highly sensitive to neutrino decays of the type $\nu_i \rightarrow \nu_j +\phi$ (especially for lifetimes in the region $\tau_\nu \sim \tU$), since it directly traces the perturbations in the present-day neutrino PSD. On the contrary, the CMB TT spectrum is affected by such late-time decays in a more indirect way, mostly through small changes in the acoustic angular scale, CMB lensing and the late integrated Sachs-Wolfe effect (LISW). In the following, we present our results for all the decay scenarios listed in \autoref{tab:decay_scenarios}, adopting the same values of $\ml$ and $\tau_\nu$ as in the background plots from \autoref{fig:PSD_A1_A2}-\autoref{fig:PSD_B1_B2}.

\begin{figure}[ht!]
\centering
\includegraphics[width=0.485\linewidth]{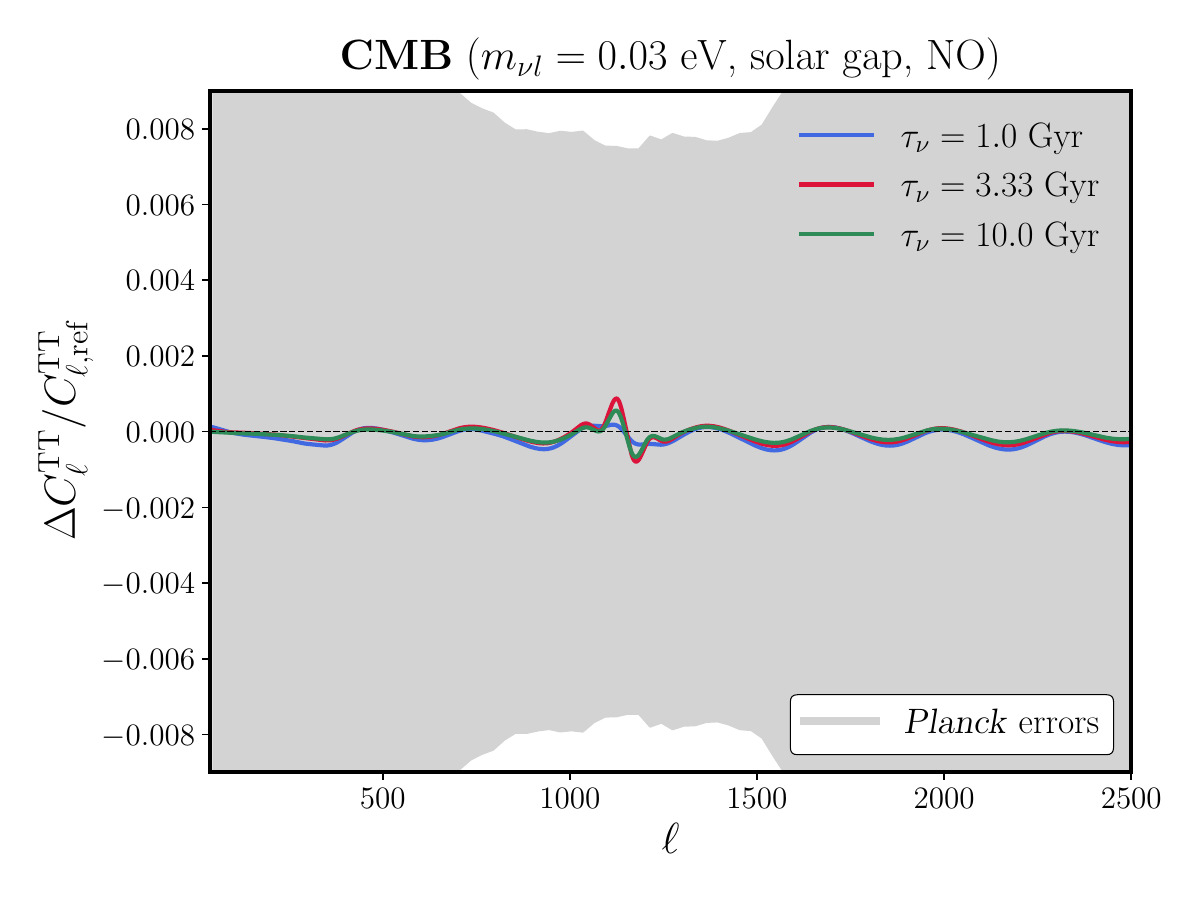} 
\includegraphics[width=0.485\linewidth]{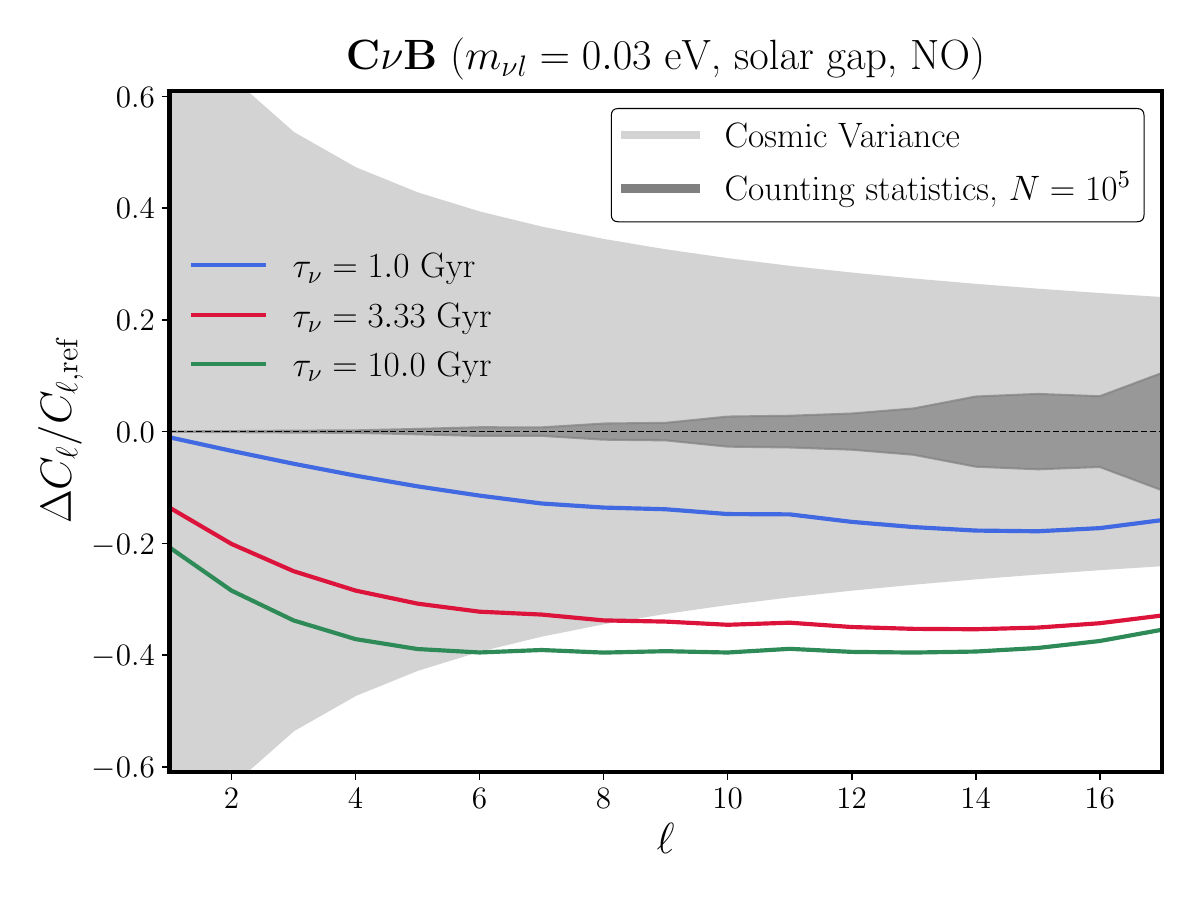} 
\caption{ \small Residuals of the CMB (\textit{left panel}) and \cnb~(\textit{right panel}) temperature power spectra in the neutrino decay scenario A2, for several neutrino lifetimes $\tau_\nu$ and a fixed daughter mass $\ml = 0.03 \ \rm{eV}$. All residuals are taken with respect to the stable limit ($\tau_\nu \rightarrow \infty$). Various contributions to the 1$\sigma$ fractional uncertainty are shown: \Planck~2018 measurement errors for the CMB, and cosmic variance and PTOLEMY counting statistics (assuming a total of $N=10^5$ capture events) for the \cnb. The \cnb~curves are smoothed with a Savitzky-Golay filter.}
\label{fig:residuals_A2}
\end{figure} 

\begin{figure}[ht!]
\centering
\includegraphics[width=0.485\linewidth]{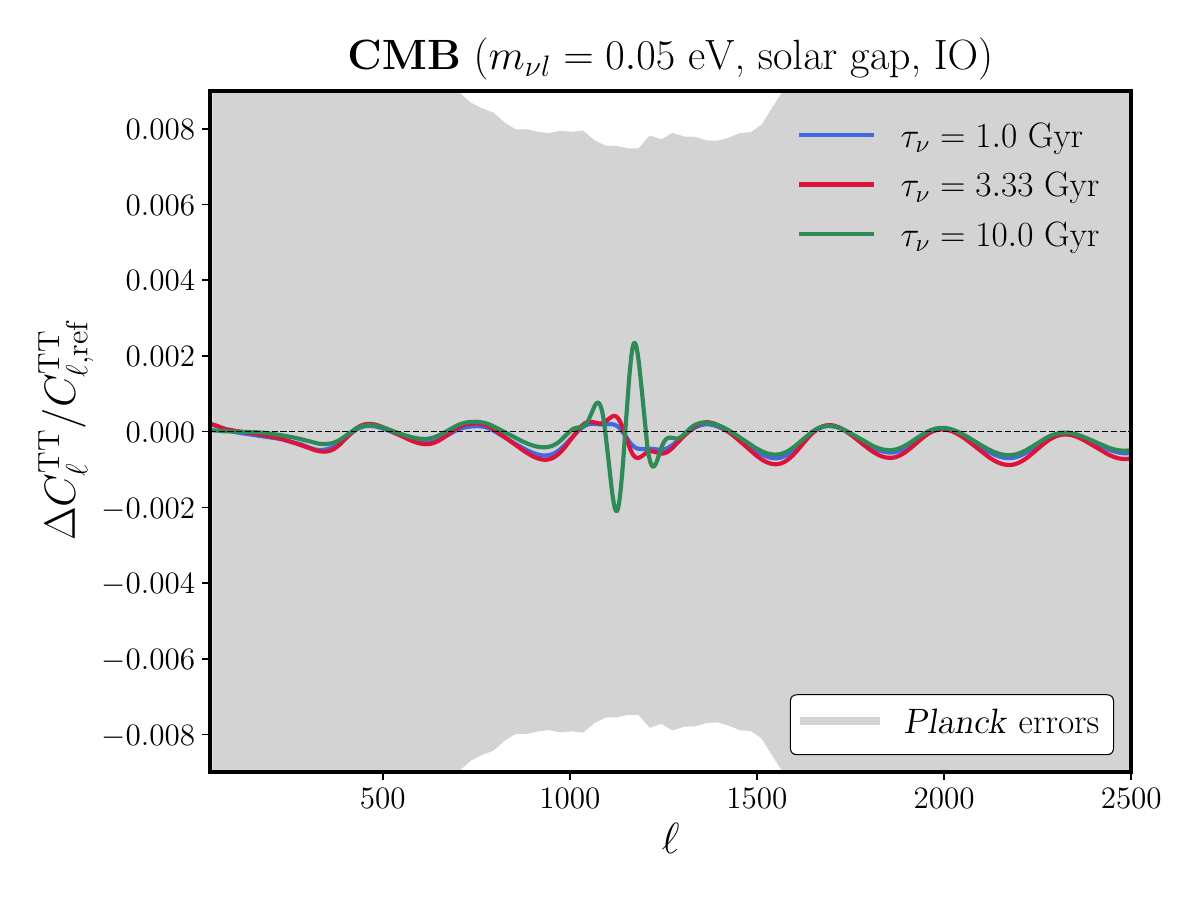} 
\includegraphics[width=0.485\linewidth]{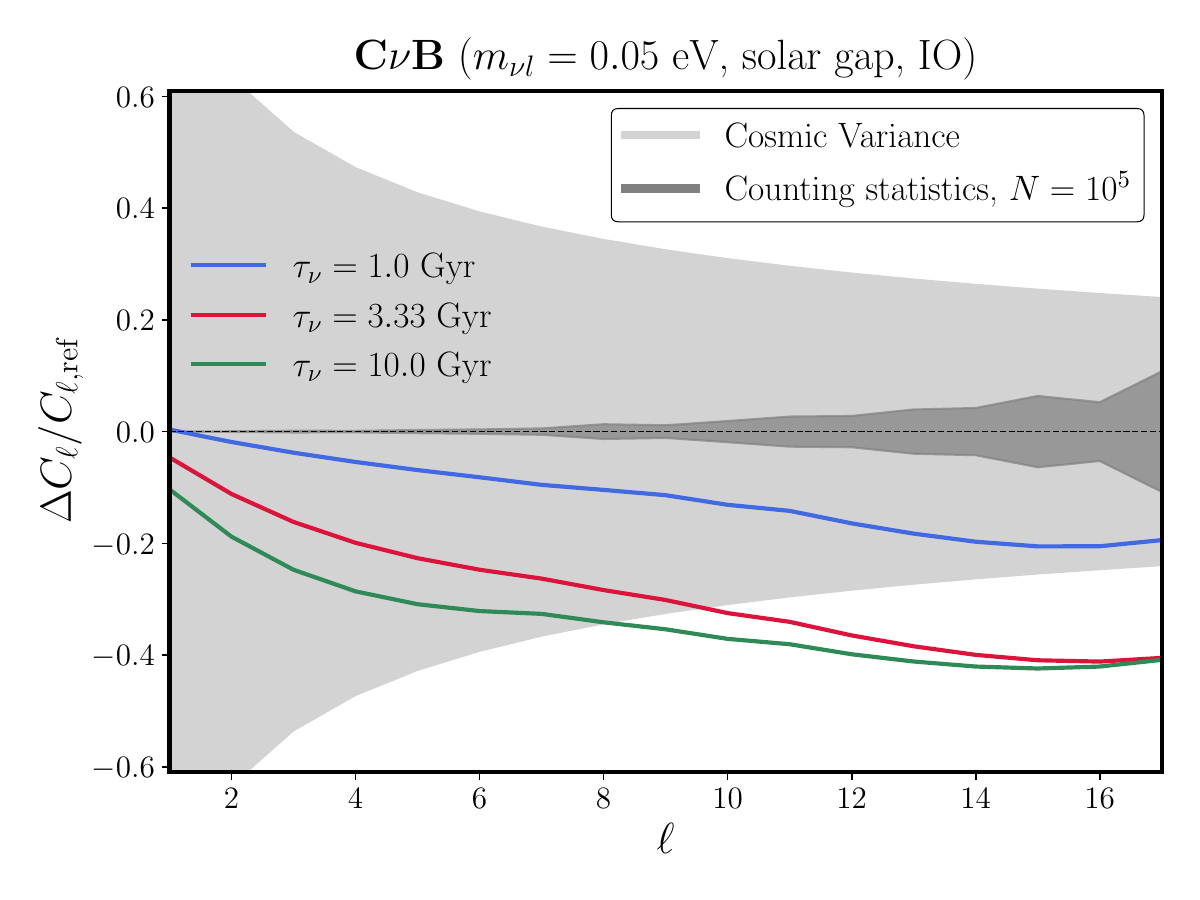} 
\caption{ \small Similar as \autoref{fig:residuals_A2}, but for the neutrino decay scenario A3 and a daughter mass $\ml = 0.05 \ \mathrm{eV}$.}
\label{fig:residuals_A3}
\end{figure}

\subsection{Decays with the solar gap}

In \autoref{fig:residuals_A2} and \autoref{fig:residuals_A3} we compare the CMB and \cnb~residuals for two distinct cases of decay scenario A (i.e., single-decay channel with the solar mass gap), assuming a fixed daughter mass $\ml$ and several values of neutrino lifetime $\tau_\nu$. On the CMB side, the effects due to neutrino decay are $\lesssim 0.2\%$, well within \Planck~1$\sigma$ uncertainties (in fact, well within cosmic variance), and are thus unobservable in any current and future CMB experiment. On the \cnb~side, the situation is markedly different: neutrino decays produce a
step-like suppression\footnote{This suppression would reflect a reduction in the amplitude of variations in neutrino capture events across the sky, but it does not imply that the isotropic component of the capture rate is smaller. In fact, the isotropic component is larger than in the stable case, since the decays increase the number density of the daughter neutrino.}, more and more pronounced as lifetime increases, which can reach $\sim 40\%$ for $\tau_\nu =10 \ \rm{Gyr}$ and exceed the cosmic variance and counting statistics uncertainties. Consequently, a future run of the PTOLEMY experiment with sufficient counting statistics (e.g. using the amplification mechanism proposed in Ref. \cite{Tully:2022erg}) could observe the impact of neutrino decays on the \cnb~anisotropies. \ 

One can also see that the residuals are very similar between scenarios A2 (NO) and A3 (IO). In this context, the main difference arises from the distinct $\ml$ values (we recall that in the IO case, $\ml$ must satisfy $\ml \geq 0.05~\mathrm{eV}$, whereas in the NO case it can take lower values).

\begin{figure}[ht!]
\centering
\includegraphics[width=0.485\linewidth]{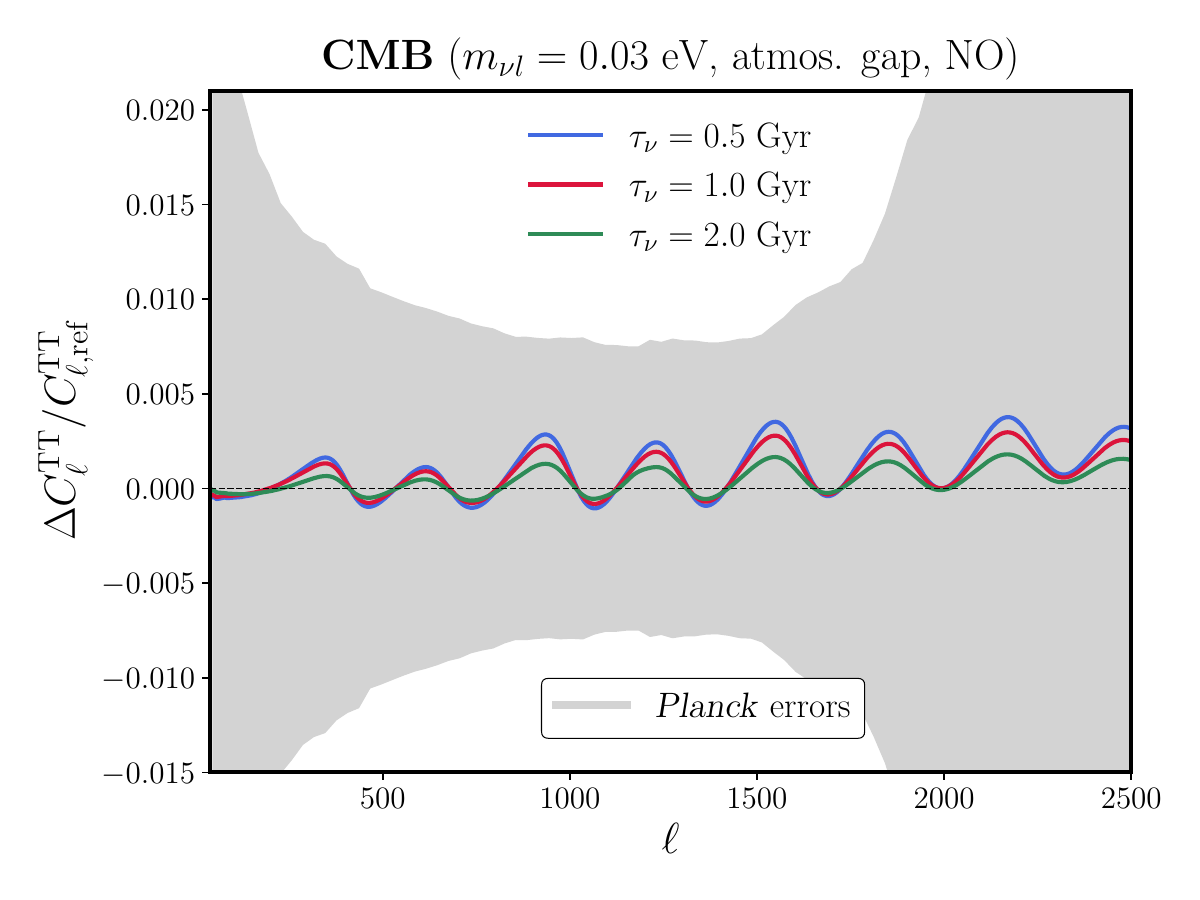 } 
\includegraphics[width=0.485\linewidth]{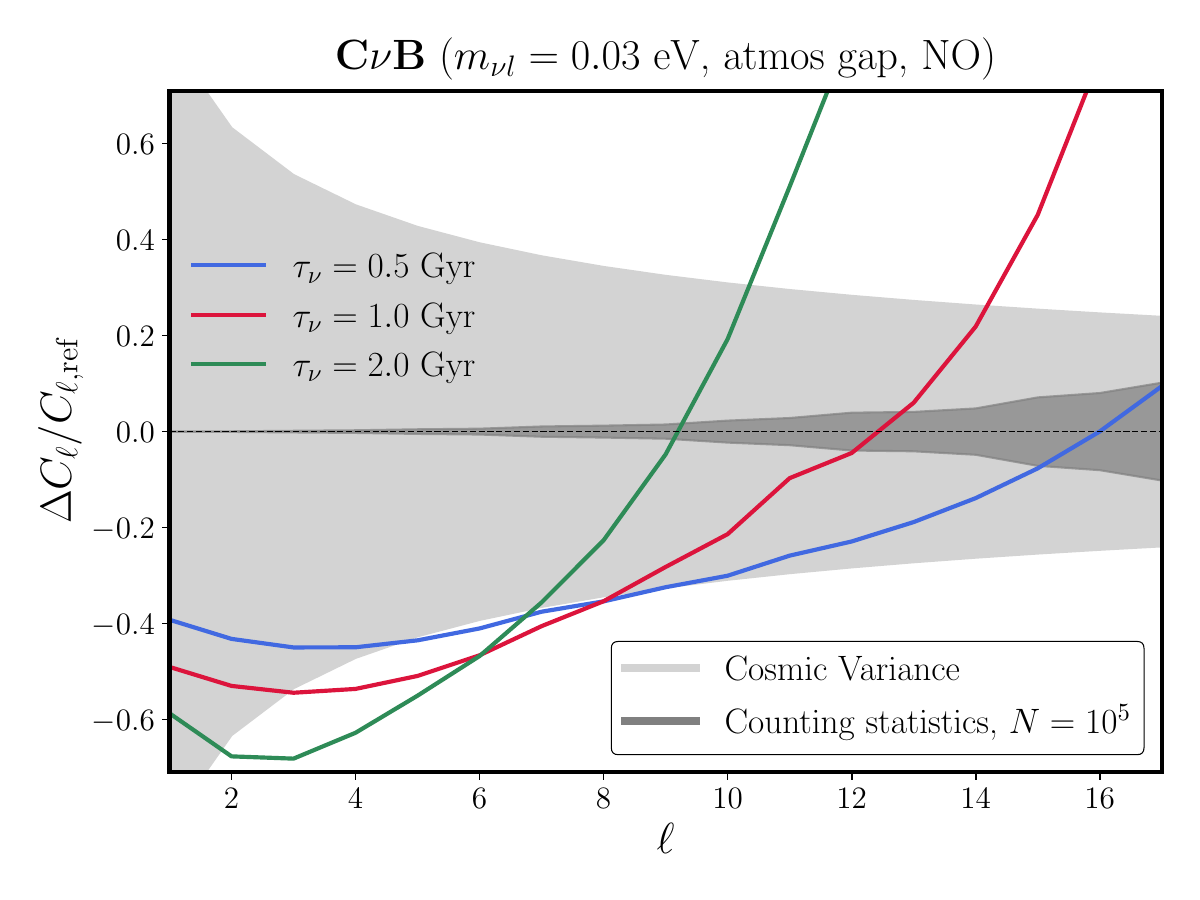 } 
\caption{ \small Similar as \autoref{fig:residuals_A2}, but for the neutrino decay scenario B1 and a daughter mass $\ml = 0.03 \ \mathrm{eV}$.}
\label{fig:residuals_B1}
\end{figure}

\begin{figure}[ht!]
\centering
\includegraphics[width=0.485\linewidth]{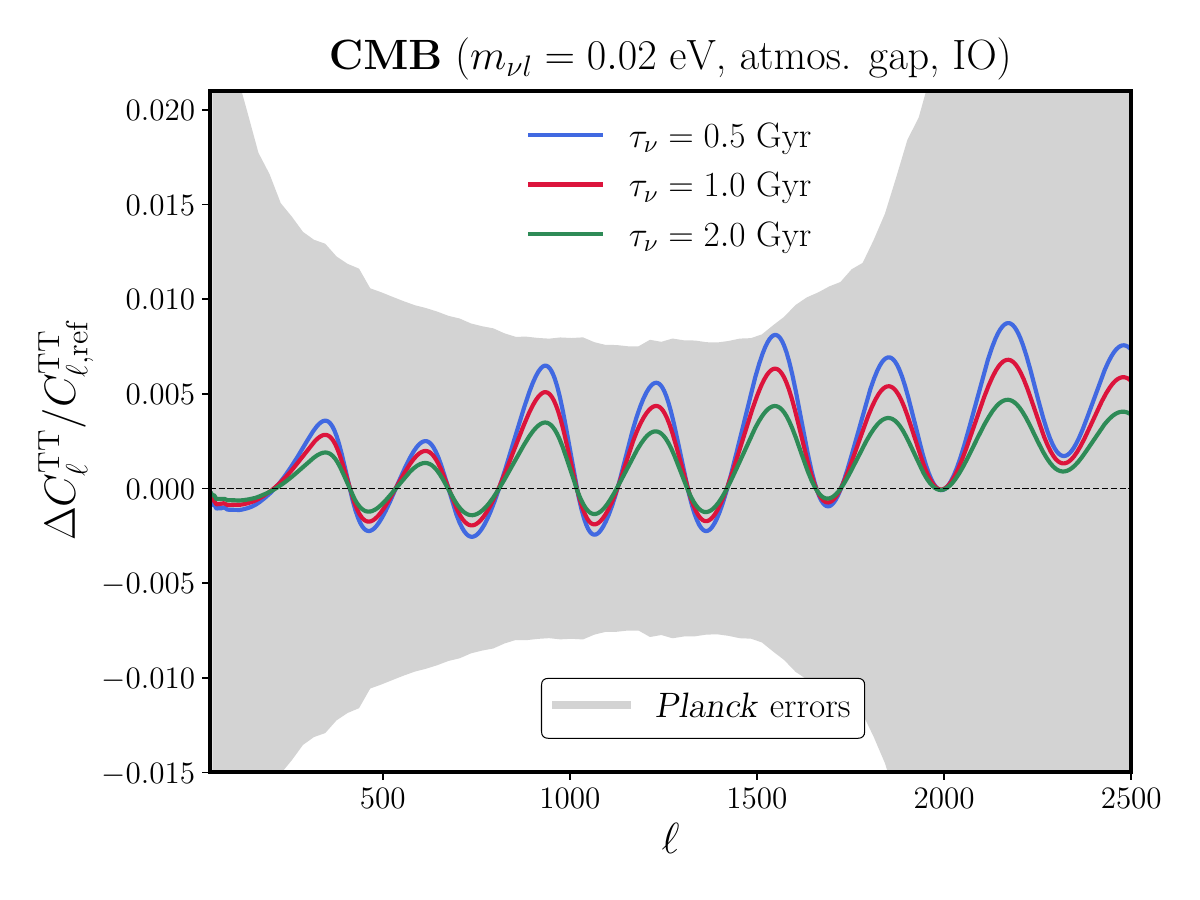 } 
\includegraphics[width=0.485\linewidth]{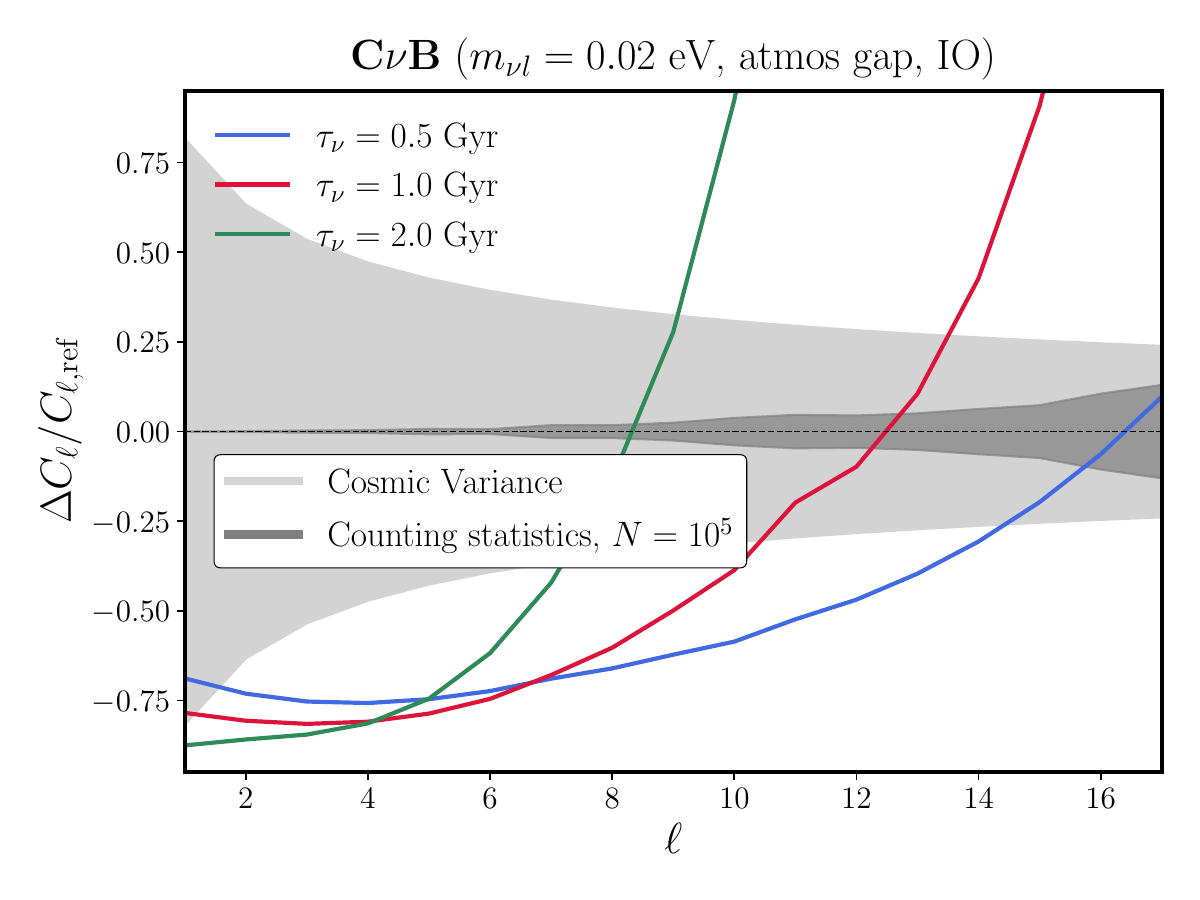 } 
\caption{ \small Similar as \autoref{fig:residuals_A2}, but for the neutrino decay scenario B2 and a daughter mass $\ml = 0.02 \ \mathrm{eV}$.}
\label{fig:residuals_B2}
\end{figure} 

\subsection{Decays with the atmospheric gap}

In \autoref{fig:residuals_B1} and \autoref{fig:residuals_B2} we compare the CMB and \cnb~residuals for two distinct cases of decay scenario B (i.e., two-decay channel with the atmospheric mass gap in the two-state approximation), assuming a fixed daughter mass $\ml$ and several values of neutrino lifetime $\tau_\nu$. At the CMB level, the signatures of neutrino decay are somewhat larger than for scenarios A, reaching up to $0.3-0.7\%$, but still remain below \Planck~uncertainties. At the \cnb~level, the effects are considerably stronger (of order $40-80\%$) and become more pronounced for longer lifetimes, though the behavior differs from that in scenarios A. In particular, neutrino decays produce a low-$\ell$ suppression and a high-$\ell$ enhancement, both of which can surpass the cosmic variance and counting statistics uncertainties for lifetimes $\tau_\nu \gtrsim 1~ \mathrm{Gyr}$. Hence, these decay scenarios could also be potentially detected or ruled out by PTOLEMY in the future.\  

In addition, we observe that scenario B2 (IO) has a stronger impact on the \cnb~spectrum than scenario B1 (NO). Indeed, under the two-state approximation, scenario B2 involves two parent neutrinos decaying into a single daughter neutrino, whereas in scenario B1 one parent neutrino decays into two daughter neutrinos. This results in a larger distortion of the daughter PSD for scenario B2 (see \autoref{fig:PSD_B1_B2}), and  consequently to a larger imprint on the \cnb~anisotropies.

\section{Conclusions and outlook} 
\label{sec:conclusion}

The angular distribution of neutrino anisotropies provides a valuable probe of LSS and fundamental neutrino properties. In this paper, we have investigated the impact of neutrino decays on the \cnb~anisotropies, within a framework where neutrinos decay non-relativistically into lighter neutrinos and a massless scalar. For this purpose, we have performed the first implementation of late-time decays $\nu_H \rightarrow \nu_l +\phi$ in a linear Einstein-Boltzmann solver, and computed the \cnb~angular power spectrum of the daughter neutrino from the Boltzmann hierarchy solutions, exploring a range of neutrino lifetimes and decay channels. 

We have found that invisible neutrino decays leave considerable imprints on the \cnb~power spectrum---about two orders of magnitude larger than on the CMB TT power spectrum---particularly for lifetimes comparable to the age of the Universe. As a consequence, neutrino decay models yielding per-mille-level (and thus undetectable) effects in CMB observables still produce signatures at the 40-80\% level in the \cnb~anisotropies, exceeding cosmic variance even for multipoles $\ell \leq 17$. A future polarized tritium target run of the PTOLEMY experiment, with sufficient counting statistics to measure $C_\ell$ up to $\ell \sim 15$, could therefore test these neutrino decays. 
We also found that the decay mass gap dictates the shape of the $C_\ell$ residuals: single-decay channels with the solar gap produce a step-like suppression, whereas two-decay channels with one near-common atmospheric gap yield a low-$\ell$ suppression and a high-$\ell$ enhancement. Hence, a \cnb~anisotropy measurement could allow to disentangle between different decay scenarios. \

Our work demonstrates the great potential of the \cnb~anisotropies to test non-standard neutrino interactions, specially if these produce significant distortions in the neutrino PSD (to which the CMB is largely insensitive \cite{Alvey:2021sji}). We hope that this gives further motivation to develop experimental techniques that could measure the multipoles in the relic neutrino sky map, like the one proposed in Ref. \cite{Tully:2022erg}. Additionally, our new \texttt{CLASS}~version will allow to confront this decaying neutrino model against the most up-to-date cosmological data, and to possibly test different models of warm dark matter decay with massless and massive decay products. \ 

For this proof-of-principle, we have made several simplifying assumptions, which should be refined as we move closer to the first direct \cnb~detection. In particular, to better connect with what PTOLEMY will observe, the temperature maps should be converted into capture rate maps, which for a given mass eigenstate depend on the fraction of electron-flavor as well as the Dirac or Majorana nature of the neutrino \cite{Tully:2021key}. Beyond this  particle-physics dependence, the translation between temperature and capture-rate maps is further complicated by astrophysical effects. Namely, Earth’s peculiar motion introduces a modulation of the signal \cite{Lisanti:2014pqa,Safdi:2014rza,Huang:2016qmh,Zimmer:2025ohu}, while the local gravitational environment---including the Milky Way halo and the nearby LSS---can significantly distort the local neutrino PSD \cite{Ringwald:2004np,Zhang:2017ljh,deSalas:2017wtt,Mertsch:2019qjv, Holm:2023rml, Holm:2024zpr, Elbers:2023mdr, Zimmer:2023jbb}. The latter is especially relevant for neutrino masses $m_\nu > 0.1 \ \rm{eV}$ \cite{Hannestad:2009xu}, where neutrino clustering dominates the signal. In this context, Ref. \cite{Zimmer:2024max} recently proposed a hybrid framework that combines perturbation theory with numerical simulations to provide a consistent description of the \cnb~anisotropies in the linear and non-linear regimes. It would be very interesting to extend such a framework to account for neutrino decays.  Finally, for a full data analysis, one would need to optimize the number of bins on the sky map to balance angular resolution against counting statistics uncertainties, and carefully model and unfold the angular smearing introduced by the polarized tritium target.  
\newpage

\acknowledgments

GFA acknowledges support from the European Research Council (ERC) under the European Union's Horizon 2020 research and innovation programme (Grant agreement No. 864035 - Undark). 
FZ and SA are part of the project ``One second after the Big Bang'' NWA.1292.19.231 which is financed by the Dutch Research Council (NWO).
SA was partly supported by MEXT KAKENHI Grant Numbers, JP20H05850, JP20H05861, and JP24K07039.

\appendix
\section{Momentum integration of the background collision term} \label{sec:app_a}

In this appendix, we derive \autoref{eq:eom_density_dr} by
carrying out the momentum integration of the background collision term of the DR. As we mentioned in section \autoref{sec:back_eqs}, we follow the steps outlined in App. A of Ref. \cite{Holm:2022eqq}, but considering a finite $\ml$ and using $\bar{f}_{\dr} \equiv \frac{1}{2} \bar{f}_{\phi}$. The starting point is the zeroth-order Boltzmann equation for $\bar{f}_{\dr} (q_3)$,
\begin{equation}
\frac{\partial \bar{f}_{\dr} \left(q_3\right)}{\partial \tau} =\frac{a^2 \mH^3 \Gamma}{\left(\mH^2 - \ml^2\right) q_3^2} \int_{q_{1-}^{\phi}}^{\infty} dq_1 \frac{q_1}{\epsilon_1} \bar{f}_{\nu H} \left(q_1\right),
\label{eq:BE_back_phi}
\end{equation}
where $q_{1-}^{\phi}$ is given in \autoref{eq:q1_phi}. We integrate \autoref{eq:BE_back_phi} over $4 \pi a^{-4} d q_3 q_3^3$, which gives $\dot{\rho}_{\mathrm{dr}}+4 a H \rho_{\mathrm{dr}}$ on the left hand side. On the right hand side, we have
\begin{equation}
    \frac{4 \pi \mH^3 \Gamma}{a^2\left(\mH^2-\ml^2\right)} \int_0^{\infty} dq_3 q_3  \int_{q_{1-}^{(\phi)}}^{q_{1+}^{(\phi)}} dq_1 \frac{q_1}{\epsilon_1} \bar{f}_{\nu H} \left(q_1\right),
\end{equation}
where $q_{1+}^{(\phi)} = \infty$. Due to the $q_3$ dependence of $q_{1-}^{(\phi)}$, we cannot evaluate the $dq_3$ integral directly.  Nevertheless, we can relax the lower bound to zero by introducing the step functions,
\begin{equation}
    \frac{4 \pi \mH^3 \Gamma}{a^2\left(\mH^2-\ml^2\right)} \int_0^{\infty} dq_3 q_3  \int_0^{\infty} dq_1 \Theta\left(q_1-q_{1-}^{(\phi)}\right) \Theta\left(q_{1+}^{(\phi)}-q_1\right) \frac{q_1}{\epsilon_1} \bar{f}_{\nu H} (q_1)  .
\end{equation}
The aim is to turn the step functions that bound the $q_1$ integral into bounds on the $q_3$ integral, which we can easily evaluate. To do so, we will use equation (A.27) of Ref.~\cite{Barenboim:2020vrr} 
\begin{equation}
    \Theta\left(1-\cos ^2 \beta^*\right) =\Theta\left(q_1-q_{1-}^{(\phi)}\right) \Theta\left(q_{1+}^{(\phi)}-q_1\right),
\label{eq:theta_beta}
\end{equation}
where $\cos \beta^*$ is given by
\begin{equation}
 \cos \beta^* =\frac{2 \epsilon_1 q_3-a^2\left(\mH^2-\ml^2\right)}{2 q_1 q_3}.  
\end{equation}
By taking the $q_3$-roots of $1-\cos ^2 \beta^*$, 
\begin{equation}
    q_{3 \pm}^{(\nu H)} = \frac{\mH^2-\ml^2}{2 \mH^2}\left(\epsilon_1 \pm q_1\right),
\end{equation}
we arrive at the useful identity
\begin{equation}
     \Theta\left(q_1 - q_{1-}^{(\phi)}\right) \Theta\left(q_{1+}^{(\phi)}-q_1\right) = \Theta\left(1-\cos ^2 \beta^*\right) = \Theta\left(q_3 - q_{3-}^{(\nu H)}\right) \Theta\left(q_{3+}^{(\nu H)}-q_3\right),
\end{equation}
which allows us to swap the integration bounds. With this, the double integral becomes 
\begin{equation}
    \frac{4 \pi \mH^3 \Gamma}{a^2\left(\mH^2-\ml^2\right)} \int_0^{\infty} dq_1 \frac{q_1}{\epsilon_1} \bar{f}_{\nu H} (q_1)  \int_0^{\infty} dq_3 q_3 \Theta\left(q_3 - q_{3-}^{(\nu H)}\right) \Theta\left(q_{3+}^{(\nu H)} - q_3\right),
\label{eq:mid_point}
\end{equation}
where we inverted the order of integration as $q_{3 \pm}^{(\nu H)}$ depend on $q_1$. The step functions turn into bounds on the $dq_3$ integral, which we can readily evaluate 
\begin{equation}
    \int_{q_{3-}^{(\nu H)}}^{q_{3+}^{(\nu H)}} dq_3 q_3=\frac{\left(q_{3+}^{(\nu H)}\right)^2-\left(q_{3-}^{(\nu H)}\right)^2}{2} = \frac{\left( \mH^2 - \ml^2 \right)^2}{2 \mH^4} \epsilon_1 q_1 .
\end{equation}
Substituting this result in \autoref{eq:mid_point} gives for the right hand side
\begin{align}
\frac{4 \pi \mH^3 \Gamma}{a^2\left(\mH^2-\ml^2\right)} \int_0^{\infty} dq_1 \frac{q_1}{\epsilon_1} \bar{f}_{\nu H} (q_1) \int_{q_{3-}^{(\nu H)}}^{q_{3+}^{(\nu H)}} dq_3 q_3 &= \frac{2 \pi \Gamma\left(\mH^2-\ml^2\right)}{a^2 \mH} \int_0^{\infty} dq_1 q_1^2 \bar{f}_{\nu H} (q_1) \nonumber \\
&= \frac{2 \pi \Gamma \left(\mH^2-\ml^2\right)}{a^2 \mH} \frac{a^3}{4 \pi} n_{\nu H} \nonumber \\ 
 &= \varepsilon a \Gamma \mH n_{\nu H}
\end{align}
where we have defined $\varepsilon \equiv \frac{1}{2} \left(1-\ml^2/\mH^2\right)$. Putting the two sides back together, we obtain the background equation of motion for the DR species, 
\begin{equation}
    \dot{\rho}_{\mathrm{dr}} +4 a H \rho_{\mathrm{dr}}= \varepsilon a \Gamma \mH n_{\nu H}.
\label{eq:eom_density_dr_app}
\end{equation}

\section{Momentum integration of the perturbed collision term}
\label{sec:app_b}

In this appendix, we derive \autoref{eq:integrated_collision} by taking the momentum averaged perturbation (\autoref{eq:integrated_pert}) of the first-order collision term of the DR (\autoref{eq:C_psi_phi}). As mentioned in section \autoref{sec:pert_eqs}, we follow along the steps in App. C of Ref. \cite{Holm:2022eqq}, but considering a finite $\ml$ and using $\bar{\Psi}_{\dr} \equiv \bar{\Psi}_{\phi}$. We note that the momentum integral of the first-order collision term contains a term with $-\Psi_{\phi,\ell}(q_3)$ which exactly cancels one of the terms in $\dot{F}_{\dr, \ell}$ (see Ref. \cite{Holm:2022eqq} for details). Hence, we are left with the integral
\begin{align}
&\left(\frac{dF_{\mathrm{dr}}}{d\tau}\right)_{C, \ell}^{(1)} \equiv \frac{r_{\mathrm{dr}}}{\left(\frac{a^4 \rho_{\mathrm{dr}}}{4\pi}\right)} \int dq_3 q_2^3 \bar{f}_{\mathrm{dr}}\left(q_3\right)\left(\frac{a^2 \mH^3 \Gamma}{\left(\mH^2 - \ml^2 \right)q_3^2 \bar{f}_{\mathrm{dr}}\left(q_3\right)} \int_{q_{1-}^{(\phi)} }^{\infty} dq_1 \frac{q_1}{\epsilon_1} \bar{f}_{\nu H}\left(q_1\right) \Psi_{\nu H, \ell}\left(q_1\right) P_{\ell}\left(\cos \beta^*\right)\right) \nonumber \\
& = \frac{4 \pi a^2 \mH^3 \Gamma}{\left(\mH^2 - \ml^2 \right) \rho_{\mathrm{crit}, 0}} \int_0^{\infty} d q_3 \int_{q_{1-}^{(\phi)}}^{\infty} d q_1 \frac{q_1 q_3}{\epsilon_1} \bar{f}_{\nu H}(q_1) \Psi_{\nu H, \ell}\left(q_1\right) P_{\ell}\left(\cos \beta^*\right) = \frac{4 \pi a^2 \mH^3 \Gamma}{\left(\mH^2 - \ml^2 \right) \rho_{\mathrm{crit}, 0}} \pazocal{I},
\label{integ_coll_partial}
\end{align}
where we have used
\begin{equation}
\frac{r_{\mathrm{dr}}}{\int d q_3 q_3^3 \bar{f}_{\mathrm{dr}}}=\frac{r_{\mathrm{dr}}}{\left(\frac{a^4 \rho_{\mathrm{dr}}}{4 \pi}\right)}=\frac{4 \pi}{\rho_{\mathrm{crit}, 0}}.
\end{equation}
We now focus on the integral $\pazocal{I}$. Using \autoref{eq:theta_beta}, and remembering that $q_{1+}^{(\phi)} = + \infty$, we can relax the lower bound to zero by introducing the step function
\begin{equation}
    \pazocal{I}=\int_0^{\infty} dq_3 \int_0^{\infty} dq_1 \Theta\left(1-\cos ^2 \beta^*\right) \frac{q_1 q_3}{\epsilon_1} \bar{f}_{\nu H}\left(q_1\right) \Psi_{\nu H, \ell}\left(q_1\right) P_{\ell}\left(\cos \beta^*\right).
\end{equation}
We can perform a change of variables by defining $u \equiv \cos \beta^*$, which gives
\begin{equation}
q_3=\frac{a^2\left(\mH^2-\ml^2\right)}{2 \epsilon_1- 2 q_1 u}, \quad \quad dq_3=\frac{2 a^2 q_1\left(\mH^2-\ml^2\right)}{\left(2 \epsilon_1- 2 q_1 u\right)^2} du,
\end{equation}
and thus the integral becomes 
\begin{equation}
\pazocal{I}=\int_0^{\infty} dq_1 \int_{u_-}^{u_+} du \Theta\left(1-u^2\right) \frac{a^4 q_1^2 \left(\mH^2-\ml^2\right)^2}{4 \epsilon_1^4} \frac{\bar{f}_{\nu H}\left(q_1\right) \Psi_{\nu H, \ell}\left(q_1\right) P_{\ell}\left(u\right)}{\left(1-\frac{q_1 u}{\epsilon_1}\right)^3}.
\end{equation}
The bounds are
\begin{equation}
    u_{-}=\lim _{q_3 \rightarrow 0} u\left(q_3\right)=-\infty < -1, \quad \quad u_{+}=\lim _{q_3 \rightarrow \infty} u\left(q_3\right)=\frac{\epsilon_1}{q_1} > +1,
\end{equation}
which means that the integration limits are actually enforced by $\Theta\left(1 - u^2\right)$ at $\pm 1$. By introducing the scattering kernel 
\begin{equation}
    \pazocal{F}_{\ell}(x)=\frac{\left(1-x^2\right)^2}{2} \int_{-1}^{+1} du \frac{P_{\ell}(u)}{(1-x u)^3},
\end{equation}
we can rewrite the integral as
\begin{align}
\pazocal{I} &=\int_0^{\infty} dq_1 q_1^2 \bar{f}_{\nu H}\left(q_1\right) \Psi_{\nu H, \ell}\left(q_1\right) \frac{a^4\left(\mH^2-\ml^2\right)^2}{2 \epsilon_1^4} \frac{\pazocal{F}_{\ell}\left(\frac{q_1}{\epsilon_1}\right)}{\left(1-\frac{q_1^2}{\epsilon_1^2}\right)^2}\nonumber\\
 &=\frac{\left(\mH^2-\ml^2\right)^2}{2 \mH^4} \int_0^{\infty} dq_1 q_1^2 \bar{f}_{\nu H}\left(q_1\right) \Psi_{\nu H, \ell}\left(q_1\right) \pazocal{F}_{\ell}\left(\frac{q_1}{\epsilon_1}\right),
\end{align}
where in the second equality we have simplified the kinematical factor
\begin{equation}
    2 \epsilon_1^4 \left(1 - \frac{q_1^2}{\epsilon_1^2}\right)^2 = 2 \left(\epsilon_1^2 - q_1^2\right) = 2 a^4 \mH^4.
\end{equation}
With this, the collision term reads
\begin{equation}
    \left(\frac{dF_{\mathrm{dr}}}{d\tau}\right)_{C, \ell}^{(1)}=\frac{4 \pi a^2 \Gamma\left(\mH^2-\ml^2\right)}{2 \mH \rho_{\mathrm{crit}, 0}} \int_0^{\infty} dq_1 q_1^2 \bar{f}_{\nu H}\left(q_1\right) \Psi_{\nu H, \ell}\left(q_1\right) \pazocal{F}_{\ell}\left(\frac{q_1}{\epsilon_l}\right).
\end{equation}
Using
\begin{equation}
\dot{r}_{\mathrm{dr}} = \frac{1}{2} \left(1-\frac{\ml^2}{\mH^2}\right) \frac{r_{\mathrm{dr}} a \Gamma \mH n_{\nu H}}{\rho_{\mathrm{dr}}} = \frac{\left(\mH^2-\ml^2\right)}{2\mH^2} \frac{a^5 \Gamma \mH n_{\nu H}}{\rho_{\mathrm{crit}, 0}},
\end{equation}
we can rewrite the prefactor as
\begin{equation}
    \frac{4 \pi a^2 \Gamma\left(\mH^2-\ml^2\right)}{2 \mH \rho_{\mathrm{crit}, 0}} = \frac{\dot{r}_{\mathrm{dr}}}{\int_0^{\infty} dq_1 q_1^2 \bar{f}_{\nu H}\left(q_1\right)},
\end{equation}
which yields the following expression for the integrated collision term
\begin{equation}
    \left(\frac{dF_{\mathrm{dr}}}{d\tau}\right)_{C, \ell}^{(1)}=\dot{r}_{\mathrm{dr}} \frac{\int_0^{\infty} dq_1 q_1^2 \bar{f}_{\nu H}(q_1) \Psi_{\nu H, \ell}(q_1) \pazocal{F}_{\ell}\left(\frac{q_1}{\epsilon_1}\right)}{\int_0^{\infty} dq_1 q_1^2 \bar{f}_{\nu H}(q_1)}.
\end{equation}

\bibliographystyle{JHEP}
\bibliography{references}

\end{document}